\documentclass[12pt]{article}

\usepackage{fullpage}
\usepackage{amssymb}
\usepackage{amsmath}
\usepackage{hyperref}
\usepackage{tensor}
\usepackage{tikz}
\usepackage{tikz-3dplot}
\usepackage{tikz-cd}
\usepackage{mathtools}

\numberwithin{equation}{section}
\newcommand{\eps}{\epsilon}

\newcommand{\deps}{\delta_\epsilon}
\newcommand{\db}{\delta_B}
\newcommand{\Gammat}{\tilde{\Gamma}}
\newcommand{\G}{{\Gamma}}
\newcommand{\Gt}{\tilde{\Gamma}}
\renewcommand{\k}{\kappa}
\renewcommand{\L}{{\mathcal L}}

\renewcommand{\O}{{\mathcal O}}
\newcommand{\La}{{\Lambda}}
\newcommand{\la}{{\lambda}}

\newcommand{\Up}{{\Upsilon}}

\newcommand{\w}{{\omega}}
\renewcommand{\t}{{\tau}}

\newcommand{\halfs}{\tfrac{1}{2}}
\newcommand{\C}{\mathbb{C}}

\newcommand{\R}{\mathbb{R}}
\newcommand{\T}{\mathbb{T}}
\newcommand{\Z}{\mathbb{Z}}

\newcommand{\re}{\textrm{Re}\,}
\DeclareMathOperator{\Tr}{Tr}
\newcommand{\bra}{\langle}
\newcommand{\ket}{\rangle}
\newcommand{\nn}{\nonumber}
\newcommand{\iso}{\cong}
\newcommand{\qvec}{{\underline{q}}}
\newcommand{\wvec}{{\vec{\omega}}}

\begin{document}

\thispagestyle{empty}
\begin{flushright} \small
UUITP-36/17
\end{flushright}

\smallskip
\begin{center} \LARGE

{\bf 7D supersymmetric Yang-Mills\\  on curved manifolds} \\[12mm] \normalsize
{\bf Konstantina Polydorou$^{a}$, Andreas Roc\'en$^b$ and Maxim Zabzine$^{a}$} \\[8mm]
{\small \it ${}^a$Department of Physics and Astronomy, Uppsala University,\\
Box 516, SE-75120 Uppsala, Sweden.}\\
\medskip 
\vspace{.5cm}
{\small \it ${}^b$Department of Mathematics, Uppsala University,\\
Box 480, SE-75106 Uppsala, Sweden.}
\medskip

\end{center}

\vspace{7mm}

\begin{abstract}
\noindent
We study 7D maximally supersymmetric Yang-Mills theory on curved manifolds that admit Killing spinors. If the manifold admits at least two Killing spinors (Sasaki-Einstein manifolds) we are able to rewrite the supersymmetric theory in terms of a cohomological complex. In principle this cohomological complex makes sense for any K-contact manifold.  For the case of toric Sasaki-Einstein manifolds we derive explicitly the perturbative part of the partition function and speculate about the non-perturbative part. We also briefly discuss the case of 3-Sasaki manifolds and suggest a plausible form for the full non-perturbative answer.

\end{abstract}

\eject
\normalsize

\tableofcontents

\section{Introduction}

Starting from the work of Pestun \cite{Pestun:2007rz} localisation of supersymmetric gauge theories on compact manifolds has attracted considerable attention, see \cite{Pestun:2016zxk} for a review of the subject. Many exact results have been produced in diverse dimensions and many consistency checks have been performed. The supersymmetric gauge theories have been placed on curved compact manifolds from 2D to 7D and explicit calculations have been performed, see \cite{Pestun:2016jze} for a brief overview. Even and odd dimensions are treated differently and the structure of the answer is different for even and odd dimensions. For low dimensions (2D and 3D, and some 4D examples like $S^3 \times S^1$) the localisation results are known very explicitly and are typically given by integrals of some special functions. These results can be obtained thanks to the fact that the localisation locus has an explicit description (e.g. flat connections on compact manifolds). Once we start to move to higher dimensions (4D-7D) the answer becomes more conjectural since it is hard to describe the localisation locus explicitly in a controlled way. This is already the case for $S^4$ in Pestun's original calculation \cite{Pestun:2007rz} where only singular point-like instantons contribute. This is to be contrasted with the calculation on $\mathbb{R}^4$ or $\mathbb{R}^4\times S^1$ where we are in full control \cite{Nekrasov:2002qd, Nekrasov:2003rj}. There non-commutativity can be introduced systematically to deal with the singular configurations. For the compact examples this option is not available so far. As we consider higher dimensional 5D-7D gauge theories on compact manifolds the situation becomes more complicated. So far in 5D-7D we only have conjectural results about the full answer. Thus localisation on compact manifolds in higher dimensions poses the main challenge, both conceptual and calculational.

The goal of this paper is to study localisation of supersymmetric Yang-Mills theory on compact 7D manifolds. In 7D the theory is unique, it is the maximally supersymmetric Yang-Mills theory and thus the answer depends only on the underlying geometry of the manifold. This paper is a continuation of  previous work \cite{Minahan:2015jta} where the 7D theory on $S^7$ was discussed. Here we turn to the case of 7D compact manifolds admitting Killing spinors. We discuss the general classification and then concentrate on the case of Sasaki-Einstein manifolds. For toric Sasaki-Einstein manifolds we give an explicit answer for the perturbative partition function in terms of toric data. This analysis is very similar in spirit to the 5D case \cite{Qiu:2014oqa}. However, in 7D it is harder to conjecture the full non-perturbative answer and we will explain why. Moreover, we do not know what the corresponding UV-completion for the 7D theory is, and thus we do not know what to match the described results with. 
 
This paper is organised as follows: in section \ref{sec:7DSYM} we review the construction of 7D supersymmetric Yang-Mills theory on curved manifolds admitting Killing spinors. We discuss the geometrical classification of such manifolds and how to extend the supersymmetry off-shell. In section \ref{sec:cohcomplex} we give an overview of contact geometry and map the supersymmetry transformation to a cohomological complex. Once formulated in cohomological variables the theory is defined for any K-contact 7D manifold. Section \ref{sec:localization} presents the localisation argument applied to the 7D theory and we present the perturbative answer in terms of superdeterminants. In section \ref{sec:equations} we discuss the localisation locus equations and relevant geometrical issues. In section \ref{sec:toric-SE} we restrict our attention to the case of toric 7D Sasaki-Einstein manifolds. We present a closed formula for the perturbative partition function in terms of toric data. In section \ref{sec:3-Sasaki} we briefly discuss the case of 3-Sasaki manifolds and present some conjectures. We finish with section \ref{sec:summary} where we summarise our results and outline the main open problems for localisation of 7D Yang-Mills theory. We also provide appendices where an overview of our conventions about supersymmetry and spinors are given. We also present some basic facts about the special functions appearing in this paper.

\section{7D supersymmetric Yang-Mills}\label{sec:7DSYM}

Recently there has been renewed interests in the construction of rigid supersymmetric gauge theories on compact manifolds. The modern point of view was initiated by Festuccia and Seiberg in \cite{Festuccia:2011ws} where gauge theory is coupled to supergravity and the gravitational degrees of freedom are treated as a background. The topological twist introduced by Witten in \cite{Witten:1988ze} is just a special case of these general considerations. Thus if the Festuccia-Seiberg program is pushed to the end then one can classify the rigid supersymmetric theories on compact manifolds and identify the relevant geometrical data, see for example \cite{Dumitrescu:2012ha}. However, in higher dimensions these considerations become increasingly complicated. Therefore in 7D we adapt a more simple-minded approach where a theory on a compact manifold is understood as a deformation of the flat theory and the corrections to the action and supersymmetry are just guessed. We do not try to solve the full classification problem for supersymmetric 7D gauge theory on compact manifolds. In our treatment we follow closely \cite{Pestun:2007rz,Minahan:2015jta} where the starting point is maximally supersymmetric Yang-Mills theory on 10-dimensional Lorentzian flat space.

\subsection{Reduction of 10-dimensional $\mathcal{N}=1$ SYM}

We start in 10-dimensional Minkowski space $\mathbb{R}^{9,1}$. Let $A_M$, $M=0,1,\dots,9$, be a gauge field, $D_M=\partial_M + A_M$ the covariant derivative and $F_{MN}=[D_M,D_N]$ the field strength. Let $\Psi_\alpha$, $\alpha=1,\dots ,16$, be a Majorana-Weyl fermion transforming in the adjoint representation of the gauge group and let $\Gamma^M$ denote the 10-dimensional Dirac matrices (see Appendix \ref{AppConv}). We will suppress spinor indices and take the trace over gauge group indices.

The 10-dimensional action \cite{BrinkSS}
\begin{equation}
\label{10daction}
S_{10}= \frac{1}{g_{10}^2}\int d^{10}x \Tr \left( \halfs F^{MN}F_{MN} - \Psi \G^M D_M \Psi \right)
\end{equation}
is invariant under the supersymmetry transformations
\begin{align} \label{susytransflat}
\delta_\epsilon A_M &= \epsilon \G_M \Psi \nn \, ,\\
\delta_\epsilon \Psi &= \halfs \G^{MN} F_{MN} \epsilon\, .
\end{align}
Here the 10-dimensional supersymmetry parameter $\eps$ is any constant real spinor and $\G^{MN} \equiv \Gt^{[M}\G^{N]}$ (see Appendix \ref{AppConv}).

We now go to a 7-dimensional theory with Euclidean signature by a dimensional reduction of the above theory. We separate the 10-dimensional gauge fields into 7-dimensional gauge fields $A_\mu$, $\mu=1,\dots,7$ and scalars $\phi_B \equiv A_B$, $B=0,8,9$, that come from the compactified dimensions. The derivatives along the compactified dimensions vanish. Note that $\phi_0$ comes from compactifying the time-like $0$-direction and so it will have a negative kinetic term. We will make a Wick rotation to deal with this in the path integral.

Let $X$ denote the 7-dimensional space. In \cite{Minahan:2015jta} the case where $X$ is the sphere $S^7$ was studied and a supersymmetric Lagrangian was constructed. Using conformal Killing spinors on the sphere one can construct a 10-dimensional Majorana-Weyl spinor $\eps$ satisfying the generalised Killing spinor equation
\begin{equation} \label{killcond}
\nabla_\mu \eps  =  \frac{1}{2r} \Gt_\mu \La \eps\, ,
\end{equation}
where $\La = \G^8 \Gt^9 \G^0$ and $r$ is a dimensionful parameter corresponding to the size of the manifold. We will drop the $r$ as we can restore it by dimensional analysis when needed. 

In this paper we will generalise the arguments of \cite{Minahan:2015jta} to other 7-dimensional manifolds $X$, not necessarily the sphere, where spinors satisfying the above equation exist. This will be the topic of the next subsection.

Taking such a spinor $\eps$ as our supersymmetry parameter we modify the supersymmetry transformations to
\begin{align} \label{susytranson}
\delta _\eps  A_M &= \eps \G _M \Psi \nn\, ,\\
\delta _\eps \Psi &= \tfrac{1}{2} F_{MN}\G ^{MN}\eps + \tfrac{8}{7} \G^{\mu B} \phi _B \nabla _\mu \eps\, .
\end{align}

In \cite{Minahan:2015jta} an action invariant under these supersymmetry transformations was constructed step by step. Here we just state the final result:
\begin{align} \label{7Daction}
S_{7D}= \frac{1}{g_{7D}^2} \int d^{7}x \sqrt{-g} \Tr \Big( 
\halfs F^{MN}F_{MN} 
- \Psi \G^M D_M \Psi 
+8 \phi^A \phi_A
+\tfrac32 \Psi \Lambda \Psi
-2 [\phi^A,\phi^B]\phi^C \varepsilon_{ABC}
\Big)\, .
\end{align}
In the last term, $\varepsilon_{ABC}$ is the anti-symmetric symbol with $\varepsilon_{890}=+1$. 
The 7D coupling constant is related to the 10D one via $g_{7D}^2 = g_{10D}^2/V_3$, where $V_3$ is the volume of the compactified space. Note that we have an $SO(1,2)$ $R$-symmetry for the scalars $\phi$. 
Also note that $\La^T = - \La$, so the fourth term is non-trivial.

In Appendix \ref{AppSusy} we verify that the action \eqref{7Daction} is indeed invariant under the supersymmetry transformations \eqref{susytranson}.

\subsection{7D Killing spinors}\label{7dkillingspinors}
\label{sectionKilling}
The supersymmetry in the section above relied on the existence of a generalised Killing spinor $\eps$ satisfying \eqref{killcond}. It is natural to ask what 7-dimensional manifolds allow us to construct such a spinor.

We start by recalling some facts about Killing spinors. We refer the reader to \cite{Blau, BoyerGalicki, Bar} for more details.

Let $X$ be a complete $n$-dimensional Riemannian spin manifold. A spinor $\psi$ is called a \emph{Killing spinor} with \emph{Killing constant} $\alpha \in \C$ if
\begin{equation}
\label{killdef}
\nabla_X \psi = \alpha X \cdot \psi
\end{equation}
holds for all tangent vectors $X$. Here $X \cdot \psi$ denotes the Clifford product of $X$ and $\psi$. 

The existence of Killing spinors is intrinsically related to the geometry of the underlying manifold.
Firstly, the existence of a non-trivial Killing spinor implies that $X$ is an Einstein manifold with constant scalar curvature 
\begin{equation}
\label{Ralpha}
R = -4n(n-1)\alpha^2 \, .
\end{equation}

We will now restrict our attention to positive\footnote{Note that in the mathematics literature one typically uses the convention $\gamma^{\{M}\gamma^{N\}} = -g^{MN}$ for the Clifford algebra, whereas in this paper, and in most of the physics literature, the opposite sign is used. This leads to opposite signs in the above equation for the scalar curvature and disagreements about whether $\alpha$ is real or imaginary etc. To avoid such ambiguities we will follow \cite{Blau} and call a Killing spinor \emph{positive} or \emph{negative} depending on if it leads to positive or negative scalar curvature respectively. The equivalent notions in the maths literature are \emph{real} and \emph{imaginary} Killing spinors, and vice versa in the physics literature.} Killing spinors. In this case $X$ is an Einstein manifold with positive curvature, and hence compact. 

Rewriting \eqref{Ralpha} as
\begin{equation}
\alpha = \pm \frac{1}{2}i \sqrt{\frac{R}{n(n-1)}}\, ,
\end{equation}
and rescaling the metric we may assume that 
\begin{equation}
\label{alphaR}
\alpha = \pm \frac{1}{2}i \, .
\end{equation}
Following \cite{Bar}, we say that the manifold $X$ is of type $(p,q)$ if it has exactly $p$ linearly independent Killing spinors with  $\alpha = + \frac12 i$ and exactly $q$ linearly independent Killing spinors with $\alpha = - \frac12 i$.

In \cite{Bar} B\"ar completed the classification of manifolds admitting positive Killing spinors. 
In particular, if $X$ is a 7-dimensional complete simply-connected Riemannian spin manifold with a non-trivial positive Killing spinor, then there are four possibilities:
\begin{enumerate}
\item $X$ is of type $(1,0)$ and is a proper $G_2$-manifold.
\item $X$ is of type $(2,0)$ and is a Sasaki-Einstein manifold (but not 3-Sasakian).
\item $X$ is of type $(3,0)$ and is a 3-Sasakian manifold (but not $S^7$).
\item $X=S^7$, which is of type $(8,8)$.
\end{enumerate}

The converse statements also hold \cite{Bar}, e.g. if $X$ is a 7-dimensional complete simply connected Riemannian spin manifold with a Sasaki-Einstein structure but not a 3-Sasaki structure, then $X$ is of type $(2,0)$, etc.

These types of manifolds are most easily described in terms of their metric cones, $C(X)$, which we define as
\begin{equation}
C(X) = X \times \R^+\, ,
\end{equation}
with metric 
\begin{equation}
ds_{C(X)}^2 = d\tau^2 + \tau^2 ds_{X}^2\, .
\end{equation}
Here $\tau$ is the coordinate on $\R^+$ and $ds_{X}^2$ is the metric on $X$.

A manifold $X$ is said to be \emph{Sasaki-Einstein} or \emph{3-Sasakian} if its metric cone $C(X)$ is Calabi-Yau or hyperk\"ahler respectively. By a \emph{proper $G_2$-manifold} we mean a 7-dimensional manifold $X$ such that its cone $C(X)$ has holonomy group $Spin(7)$. These manifolds can also be characterised as the 7-dimensional manifolds admitting a nearly parallel $G_2$-structure. That is, they possess a 3-form $\Phi$ associated to the $G_2$-structure satisfying  $d \Phi = -8 \alpha (* \Phi)$, see \cite{Friedrich:1997}.

The existence of Killing spinors in 7D allows us to construct 10D spinors that satisfy \eqref{killcond}. We give more details about this in Appendix \ref{AppKill}.

It is interesting to contrast the 7D situation to the perhaps more well-known case of 5D. The analogous statement about the existence of Killing spinors in 5D is:
\begin{enumerate}
\item $X_5$ is of type $(1,1)$ and is a Sasaki-Einstein manifold (but not $S^5$).
\item $X_5=S^5$, which is of type $(4,4)$.
\end{enumerate}
 The 5D supersymmetric action has been constructed in \cite{Hosomichi:2012ek} and 
localisation of supersymmetric gauge theories on such 5D manifolds is discussed in the review \cite{Qiu:2016dyj} (see the references therein). Comparing the two lists above we see that we get a wider variety of geometries to work with in 7D than in 5D. Proper $G_2$ manifolds are unique to dimension seven, while 3-Sasaki manifolds require the dimension to be of the form $n=4m-1$. Note that 3-dimensional Sasaki-Einstein manifolds are automatically 3-Sasakian, so dimension seven is also the smallest dimension where the notions of Sasaki-Einstein and 3-Sasakian are distinct.

\subsection{Off-shell extension} \label{offshellextension}

We now return to the discussion of our 7D Yang-Mills theory. In order to apply a localisation argument the supersymmetry generator $\deps$ needs to square to a symmetry of the theory (such as Poincar\'e or gauge symmetries). However, as currently formulated in \eqref{susytranson} it only does so on-shell, i.e. using the equations of motion. The next step is thus to formulate the supersymmetry transformations off-shell. This was first done in \cite{Berkovits:1993hx} for 10-dimensional flat space and in \cite{Fujitsuka:2012wg} for curved backgrounds.
Note that while our manifold may admit several Killing spinors, we only take the supersymmetry off-shell for a single $\eps$ to apply the localisation procedure. In general, it is not known if an off-shell formulation for more than one supersymmetry is possible.

Following \cite{Pestun:2007rz,Minahan:2015jta} we begin by fixing one of the Killing spinors $\eps$  and picking a set of seven bosonic pure spinors $\nu_m$, $m=1,\dots,7$ satisfying the relations
\begin{align} \label{purespinordef}
\eps \G^M \nu_m &=0  \nn\, ,\\
\nu_m \G^M \nu_n &= \delta_{mn} v^M\, ,\\
\nu^m_\alpha \nu^m_\beta + \eps_\alpha \eps_\beta &= \halfs v^M \Gt_{M\alpha\beta} \nn\, .
\end{align}

Here $v^M$ denotes the vector field 
\begin{equation} \label{eq:vdef}
v^M = \eps \G^M \eps\, .
\end{equation}
Note that these relations only determine the $\nu$'s up to an internal $SO(7)$ symmetry.

At this point we will restrict our attention to the case when $X$ is a Sasaki-Eistein manifold\footnote{This includes 3-Sasaki and $S^7$ as subcases.}. The reason is that $v^\mu$, i.e. the vector field $v$ restricted to the 7D space, will be identically zero for the proper $G_2$ case, see appendix \ref{AppKill}. We hope to discuss the proper $G_2$ case in future work.

For the Sasaki-Einstein case we are free to choose $v^0=1$ and $v^8=v^9=0$. Using the triality identity \eqref{triality} it then follows that $v^\mu v_\mu=1$.

For each pure spinor we introduce an auxiliary field $K^m$ which transforms non-trivially under the supersymmetry. The supersymmetry transformations are modified to
\begin{align} \label{susytransoff}
&\deps  A_M = \eps \G _M \Psi \nn\, ,\\
&\deps \Psi = \halfs F_{MN}\G ^{MN}\eps + \tfrac{8}{7} \G^{\mu B} \phi _B \nabla _\mu \eps + K^m \nu_m\, , \\
&\deps K^m =  -\nu^m \G^M D_M \Psi + \tfrac32 \nu^m \La \Psi \nn\, .
\end{align}

It can be shown, see \cite{Pestun:2007rz,Minahan:2015jta} or Appendix \ref{offshell}, that this off-shell supersymmetry squares to symmetries of the theory. Schematically,
\begin{equation}
\deps^2 = -L - G - R - S\, ,
\end{equation}
where $L$ is a Lie derivative along the vector field $v$, $G$ is a gauge transformation, $R$ is the $R$-symmetry and $S$ is the $SO(7)$ rotations of the auxiliary fields $K^m$.

To make the action invariant under the above supersymmetry transformations the following free quadratic term is added to \eqref{7Daction}
\begin{equation}
S_{aux} = - \frac{1}{g_{7}^2} \int d^7x \sqrt{-g} \Tr K^m K_m \, .
\end{equation}
Note that this term is invariant under the internal $SO(7)$ symmetry. Also due to the minus sign in front, $K$ needs to be Wick-rotated in the path integral. The full form of the off-shell supersymmetric Lagrangian is thus $S=S_{7D}+S_{aux}$,
\begin{align} \label{7DactionOffShell}
S= \frac{1}{g_{7D}^2} \int d^{7}x \sqrt{-g} \Tr \Big( 
\halfs F^{MN}F_{MN} 
- \Psi \G^M D_M \Psi 
&+8 \phi^A \phi_A
+\tfrac32 \Psi \Lambda \Psi \notag \\
&-2 [\phi^A,\phi^B]\phi^C \varepsilon_{ABC}
- K^m K_m
\Big)\, .
\end{align}

\section{Cohomological complex and contact geometry}\label{sec:cohcomplex}

In this section we map the fields of our theory to a set of differential forms and turn the supersymmetry transformations into a cohomological complex. This will cast the theory in the geometrical language needed to apply the localisation technique. Although we restricted our attention to Sasaki-Einstein manifolds in the last section, the complex we find here is in principle defined for any K-contact manifold.

\subsection{Contact geometry}
We start by briefly reviewing some facts from contact geometry, see e.g. \cite{Blair2010,Geiges} for more details.

A contact structure on a $(2n+1)$-dimensional smooth manifold can be described in terms of a contact 1-form $\kappa$ such that $\k \wedge (d\k)^n \neq 0$. This gives rise to a hyperplane field $\ker \k$ which we will refer to as the horizontal space. On this horizontal space $d\k$ is non-degenerate and acts like a symplectic form. Associated to a contact form $\k$ there exists a unique Reeb vector field $R$ satisfying
\begin{align}
\iota_R d\k = 0\, ,\label{reeb1} \\
\iota_R \k = 1 \label{reeb2}\, .
\end{align}
 
Using the projectors 
\begin{equation}
P_V=\k \wedge \iota_R \quad \text{ and } \quad  P_H = 1-P_V \, , \label{eq:HVProjDef}
\end{equation}
we can decompose differential forms into vertical and horizontal parts
\begin{equation}
\Omega^k = \Omega^k_V \oplus \Omega^k_H \, .
\end{equation}

Given a contact structure one can always find a Riemannian metric $g$ and a $(1,1)$-tensor field $J$ such that
\begin{align}
&J^2 = -I + \k \otimes R  \label{eq:cmsJ2}\, , \\
&g(JX,JY)  = g(X,Y) - \k(X)\k(Y)  \label{eq:cmsgk}\, ,\\
&d\k(X,Y) = g(X,J Y) \label{eq:cmsdk}\, .
\end{align}
We then say that we have a contact metric structure.

As a consequence we have
\begin{align}
JR &=0 \, ,\\
\k \circ J &=0 \, ,\\
\k(X) &= g(X,R)\, .
\end{align}

For such a metric we have the useful identity
\begin{equation} \label{ivstar}
i_R (* \alpha_p)= (-1)^p * (\k \wedge \alpha_p) \, ,
\end{equation}
where $\alpha_p$ is a $p$-form and $*$ denotes the Hodge star.

We also have the following relation between volume forms
\begin{equation} \label{eq:volrel}
vol_g = \frac{(-1)^n}{2^n n!} \k \wedge (d\k)^n \, .
\end{equation}

If in addition $R$ is Killing with respect to the metric,
\begin{equation}
\L_R g = 0\, ,
\end{equation}
then we say that we have a K-contact structure.

On the horizontal space $J$ acts as an almost complex structure under which the horizontal forms decompose as
\begin{equation}\label{decomp-11}
\Omega_H^k = \bigoplus_{p+q=k} \Omega_H^{(p,q)}\, .
\end{equation}

The K-contact condition implies that this decomposition is preserved by $\L_R$, i.e. $\L_R J =0$.

Now let us focus on dimension seven.

We already saw that differential forms can be decomposed into vertical and horizontal parts via \eqref{eq:HVProjDef}
\begin{equation}
\alpha = \alpha_V + \alpha_H\, .
\end{equation}
We will now turn our attention to 2-forms and 3-forms in 7D and consider further decompositions of such forms.

On 2-forms we define the projectors\footnote{The numerical factor is due to our conventions where $(d\k)_{\mu\nu}(d\k)^{\mu\nu}=24.$}
\begin{equation} \label{eq:hatOpDef}
\check{P} = \tfrac{1}{12} \left[* (\cdot \wedge * d\k)\right] d\k
\quad \text{ and } \quad  \hat{P} = 1- \check{P}\, ,
\end{equation}
and get the decomposition
\begin{equation} \label{eq:hatDecompDef}
\alpha = \hat{\alpha} + \check{\alpha} = \hat{\alpha} + \tfrac{1}{24} \tilde{\alpha} d\k\, ,
\end{equation}
where $\tilde{\alpha}  = \alpha_{\mu\nu}(d\k)^{\mu\nu}$ and $\hat{\alpha}_{\mu\nu}(d\k)^{\mu\nu}=0$.

We can think of $\check{P}$ as picking up the part of $\alpha$ proportional to $d\k$. Thus by \eqref{reeb1} $\check{\alpha}$ is necessarily horizontal while $\hat{\alpha}$ may have both a vertical and a horizontal component.

On the horizontal part of $\hat{\alpha}$ the operator 
$\iota_R *(d\k \wedge \cdot)$ squares to $4$ and we can define the projectors 
\begin{equation}\label{eq:pmDecompDef}
P^\pm = \tfrac12 (1 \pm \tfrac{1}{2} \iota_R * (d\k \wedge \cdot))\, ,
\end{equation}
giving the decomposition
\begin{equation}
\hat{\alpha}_H = \hat{\alpha}_H^+ + \hat{\alpha}_H^-\, ,
\quad \text{ where } \quad  
\iota_R * (d\k \wedge \hat{\alpha}_H^\pm) = \pm 2 \hat{\alpha}_H^\pm\, .
\end{equation}
One can also check that
\begin{equation}
\iota_R * (d\k \wedge d\k) = -4 d\k\, .
\end{equation}

We thus have the following decomposition of 2-forms
\begin{align} \label{eq:2FormDecomp}
\Omega^2 &= \Omega^2_V \oplus \hat{\Omega}^{2+}_H \oplus \hat{\Omega}^{2-}_H \oplus \check{\Omega}^2_H\, .
\end{align}
These spaces can also be characterised as the $0, +2, -2$ and $-4$ eigenspaces of the operator $\iota_R * (d\k \wedge \cdot)$.

One can check that these spaces are mutually orthogonal with respect to the standard inner product on forms
\begin{equation}
(\alpha,\beta) = \int \alpha \wedge * \beta\, .
\end{equation}

In summary, we have the following orthogonal decomposition of a 2-form $\alpha$:
\begin{align}
\alpha &= \alpha_V + \hat{\alpha}_H^+ + \hat{\alpha}_H^-  + \check{\alpha}_H \nn\\
&= \alpha_V + \hat{\alpha}_H^+ + \hat{\alpha}_H^-  + \tfrac{1}{24} \tilde{\alpha}d\k \label{eq:2FormDecomp5} \, ,
\end{align}
where in the second line we wrote out the $d\k$ component explicitly as in \eqref{eq:hatDecompDef}. The decompositions \eqref{decomp-11} and \eqref{eq:2FormDecomp} are related to each other through the following relations:
\begin{equation}
\hat{\Omega}^{2-}_H  = \Omega_H^{(2,0)} \oplus \Omega_H^{(0,2)} \quad \text{and} \quad
\hat{\Omega}^{2+}_H \oplus  \check{\Omega}^2_H = \Omega_H^{(1,1)}\, ,
\end{equation}
where $\hat{\Omega}^{2-}_H$ is a six dimensional space and $\Omega_H^{(1,1)}$ is nine dimensional. In what follows we will use the short hand notation $\check{\Omega}^2_H = \Omega^0 d\k$ for the one-dimensional space proportional to $d\k$.

Let us now turn to 3-forms. These can of course also be decomposed into vertical and horizontal components, $\beta = \beta_V + \beta_H$.
On the horizontal part the operator $\iota_R * $ squares to $-1$ and we decompose it into $\pm i$ eigenspaces via the projectors
\begin{equation}
\mathcal{P}^\pm = \tfrac12 (1 \mp  i \iota_R *)\, .
\end{equation}
We thus decompose a 3-form as
\begin{equation}\label{eq:3FormDecomp}
\beta = \beta_V + \beta_H^+ + \beta_H^-\, ,
\quad \text{ where } \quad  
\iota_R * (\beta_H^\pm) = \pm i \beta_H^\pm\, .
\end{equation}

\subsection{Cohomological complex}

We now proceed to map our fields to differential forms and write down the cohomological complex corresponding to the supersymmetry transformations \eqref{susytransoff}.

In the previous section we introduced the vector field $v^M = \eps \Gamma ^M \eps$. For Sasaki-Einstein manifolds we had $v^8=v^9=0, v^0=1$ and $v^\mu v_\mu=1$. Sasaki-Einstein manifolds are contact manifolds and we identify $R^\mu = v^\mu$, $\k_\mu = v_\mu = g_{\mu\nu}R^\nu$, as the Reeb and the contact form respectively. Then the property $v^\mu v_\mu=1$ corresponds to the Reeb condition \eqref{reeb2}.

Using \eqref{killcond} and the anti-symmetry of $\La$ we find
\begin{equation}
(d\k)_{\mu\nu} = 2 \nabla_{[\mu} \k_{\nu]} = -2 \eps \Gt_{\mu\nu}\La \eps\, .
\end{equation}
From this expression it is straightforward to check that $R^\mu (d\k)_{\mu\nu}=0$ which is the Reeb condition \eqref{reeb1}.

We define $J$ via \eqref{eq:cmsdk}
\begin{equation} \label{csks}
\tensor{J}{^\mu_\nu} = \tfrac12 g^{\mu \rho} (d\k)_{\rho \nu} = -\eps \tensor{\Gt}{^\mu_\nu} \La \eps\, .
\end{equation}

Using triality \eqref{triality} and \eqref{purespinordef} one can check that
\begin{equation}
\tensor{J}{^\mu _\sigma}\tensor{J}{^\sigma _\nu} = -\tensor{\delta}{^\mu_\nu} + v^\mu v_\nu\, , 
\end{equation}
from which it follows that \eqref{eq:cmsJ2} and \eqref{eq:cmsgk} hold.

From the Killing spinor equation it follows that $R$ is Killing, and thus $R,\k, J$ defined in this way gives a K-contact structure.

It will be convenient to make some field redefinitions. Following \cite{Minahan:2015jta} we decompose the fermion field $\Psi$, which has 16 independent degrees of freedom, as
\begin{align}
\Psi = \sum_{M=1}^9 \Psi_M \Gt^M \G^0 \eps + \sum_{m=1}^7 \Up_m \nu^m\, ,
\end{align}
where, by using \eqref{purespinordef}, we have
\begin{align}
\Psi_M &= \eps \G_M \Psi \, ,\\
\Up_m &= \nu_m \G^0 \Psi\, .
\end{align}
Note that the $\Psi_0$ is not independent but equal to a sum of the remaining fermionic vectors,  $\Psi_0 = \eps \G^0 \Psi = - \sum_{M=1}^9 (\eps \G^M \eps)\Psi_M=- \sum_{M=1}^9 v^M \Psi_M$.

We then introduce the field $H_m$ as the $\deps$-variation of $\Up_m$,
\begin{align}
H_m = \deps \Up_m = K_m + 4\phi_0(\nu_m \La \eps) + \halfs F_{MN}(\nu_m\G^{MN0}\eps) - 4\phi_A(\nu_m\G^0\Gt^A\La\eps)\, .
\end{align}
We can think of $H_m$ as a shift of the field $K_m$.

In terms of these fields, the supersymmetry transformations \eqref{susytransoff} read \cite{Minahan:2015jta}
\begin{align} \label{susytransoffre}
&\deps A_M = \Psi_M \, ,\\
&\deps \Psi_M = -v^N F_{NM} - [\phi_0,A_M] - 4 \phi_A(\eps \Gt_{MA}\La \eps)\, , \\
&\deps \Up_m = H_m \, ,\\
&\deps H_m = -v^\mu D_\mu \Up_m - [\phi_0,\Up_m] - (\nu_m\G^\mu \nabla_\mu \nu_n) \Up^n + \tfrac32 (\nu_m \La \nu_n)\Up^n\, . \label{susytransoffre-end}
\end{align}
where we have taken all $M,N,A \neq 0$.

We now proceed to map these fields into differential forms.

We define the fermionic 2-form $\Up$ via
\begin{equation}
\Up_{\mu\nu} = \Up_m(\nu_m\G_{\mu\nu 0}\eps)\, ,
\end{equation}
and its bosonic superpartner
\begin{equation}
H_{\mu\nu} = \deps \Up_{\mu\nu} = H_m(\nu_m\G_{\mu\nu 0}\eps)\, .
\end{equation}
It is argued in \cite{Minahan:2015jta} that $\Up$ and $H$ are horizontal and can be decomposed into $(2,0)$ and $(0,2)$-forms and a $(1,1)$-form proportional to $d\k$.

The bosonic 3-form $\Phi$ is defined through
\begin{equation}\label{Phi_def}
\Phi_{\mu \nu\lambda} = \halfs \phi_A(\eps \G_{\mu\nu\lambda} \G^{A0}\eps)\, ,
\end{equation}
and its fermionic superpartner $\eta$ via
\begin{equation}
\eta_{\mu\nu\lambda} = \deps \Phi_{\mu \nu\lambda} = \halfs \Psi_A(\eps \G_{\mu\nu\lambda} \G^{A0}\eps)\, .
\end{equation}
In \cite{Minahan:2015jta} it is shown that these are horizontal and decompose into $(3,0)$ and $(0,3)$-forms. 

Finally, we simply rename $\phi_0=\sigma$. The mappings are summarised in Table \ref{table:CohomMaps}.
\renewcommand{\arraystretch}{1.4}
\begin{table}[h!]
\begin{center}
	\begin{tabular}{|c|c|c|c|}
		\hline
		\multicolumn{2}{|c|}{Bosons}& \multicolumn{2}{|c|}{Fermions}\\
		\hline
		$A_\mu$ & $A$ ~~~{\rm connection} & $\Psi_\mu$&$\psi\in \Omega^1 $ \\ \hline
		$H_m$ & $H\in\Omega_H^{(2,0)}\oplus \Omega_H^{(0,2)}\oplus \Omega^0 d\k$ & $\Up_m$&$\Upsilon\in \Omega_H^{(2,0)}\oplus \Omega_H^{(0,2)}\oplus \Omega^0 d\k $ \\ \hline
		$\phi_8,\phi_9$ & $\Phi\in\Omega_H^{(3,0)}\oplus \Omega_H^{(0,3)} $ & $\Psi_8,\Psi_9$&$\eta\in \Omega_H^{ (3,0)}\oplus \Omega_H^{(0,3)} $ \\ \hline
		$\phi_0$ & $\sigma \in \Omega^0 $& \multicolumn{2}{|c|}{} \\ \hline
	\end{tabular}
	\caption{\textit{Mappings of the bosonic and fermionic fields of our theory to differential forms. $(X,X')$-pairs of bosons and fermions appearing in the transformations \eqref{cohomoltranf1}-\eqref{cohomoltranf2} are written on the same line. Note that we have suppressed the Lie algebra dependence.}}\label{table:CohomMaps}
\end{center}\end{table}

After Wick-rotating $\phi_0$ and $K_m$ the supersymmetry transformations \eqref{susytransoffre}-\eqref{susytransoffre-end} are mapped to the following cohomological complex  \cite{Minahan:2015jta}: 
\begin{align}\label{a-variation}
&\deps A = \psi\, , \\ \label{beg-s-variation}
&\deps \psi = -\iota_R F + i G_{\sigma}A \, ,\\\label{s-variation}
&\deps \sigma = i \iota_R \psi \, ,\\
&\deps \Phi = \eta \, ,\\
&\deps \eta = -\L_R^A \Phi + iG_{\sigma}\Phi \, , \\
&\deps \Up = H\, , \\
&\deps H = -\L_R^A \Up + iG_{\sigma}\Up  \label{end-s-variation}\, .
\end{align}
Here $d_A$ is the de Rham differential coupled to the connection $A$, $d_A = d + [A,\cdot ]$, and $\L_R^A$ is the corresponding Lie derivative along the Reeb vector field, $\L_R^A = \iota_R d_A + d_A \iota_R=\L_R + [\iota_R A, \cdot ] $. The gauge transformation $G_\sigma$ is given by $G_{\sigma} A = d_A {\sigma}$ on the gauge field and $G_{\sigma} X = -[{\sigma},X]$ on the other fields.

Redefining the field $\sigma \rightarrow -\sigma + i\iota_R A$ we can write the above transformations as
\begin{align}\label{cohomoltranf1}
&\deps X = X'\, ,\\
&\deps X' = -\L_R X -iG_{\sigma}X \, ,\\ \label{cohomoltranf2}
&\deps \sigma = 0\, ,
\end{align}
where the $(X,X')$ pairs are given by $(A,\psi)$, $(H,\Up)$ and $(\Phi,\eta)$. We see that $\deps^2 = -\L_R -iG_{\sigma}$, and thus $\deps$ squares to symmetries of the theory.

We note that the above cohomological complex \eqref{a-variation}-\eqref{end-s-variation} is defined for any 7D K-contact manifold. The map between the supersymmetric variables and differential forms (together with their transformations) is invertible for the case of Sasaki-Einstein manifolds. This can be shown using identities in Appendix \ref{AppConv}. The Sasaki-Einstein geometry corresponds to a concrete choice of Reeb vector field $R$ and we will think of this as the `unsquashed' geometry. In the localisation calculation we will use the cohomological complex and  allow $R$ to be an arbitrary combination of isometries. The underlying geometry will then be Sasaki (not Sasaki-Einstein), and we refer to this as the `squashed' geometry. We strongly believe that there is a modification of the Killing spinor equation and supersymmetry transformation to accommodate Sasaki geometry, but we do not know how to analyse this at the moment.

\section{Localisation of 7D theory}\label{sec:localization}

In order to fix notation, we briefly recall the localisation procedure for supersymmetric gauge theories, see \cite{Pestun:2016jze} for a review. 

Consider a theory with partition function given by the path integral
\begin{equation}
Z = \int \mathcal{D}\Phi \, e^{-S}\, ,
\end{equation}
where $S$ is the action and $\mathcal{D}\Phi$ the integration measure for all fields. Assume we have a fermionic symmetry, generated by $Q$, that leaves both the action and measure invariant. Then we can deform the path integral by adding a $Q$-exact term:
\begin{equation} \label{pideformed}
Z[t] = \int \mathcal{D}\Phi \, e^{-S-tQV}\, ,
\end{equation}
where $t$ is some parameter. If additionally $Q^2V=0$ then $Z[t]$ is in fact independent of $t$ and we can evaluate it in the limit $t\rightarrow \infty$. In this limit, the path integral localises to the zeros of $QV$, called the fixed point locus. The integral then reduces to the action evaluated at the fixed point locus and determinant factors from the one-loop approximation (which becomes exact in this limit).

\subsection{Fixed point locus}

The potential $V$ in \eqref{pideformed} is chosen to be \cite{Pestun:2007rz,Minahan:2015jta} 
\begin{equation}
V = \int dx^7 \sqrt{-g} \Tr \left( \Psi \overline{\deps \Psi} \right) \label{bsrt-term}\, ,
\end{equation}
where
\begin{equation}
\overline{\deps \Psi} = \halfs F_{MN}\Gt^{MN} \G^0 \eps + \tfrac{8}{7} \Gt^{\mu B} \phi _B \G^0  \nabla _\mu \eps - K^m \G^0 \nu_m
\end{equation}
is the conjugate of the transformation $\deps \Psi$ in \eqref{susytransoff}.

We now want to find the fixed point locus, i.e. the zeros of $\deps V$.

The bosonic part of $\deps V$ is given by
\begin{align}
\deps V |_{bos} =  \int dx^7 \sqrt{-g} \Tr \left( \deps \Psi \overline{\deps \Psi} \right)\, .
\end{align}

By expanding the integrand $\deps \Psi \overline{\deps \Psi}$ and massaging the terms into a sum of squares, see \cite{Pestun:2007rz,Minahan:2015jta}, we find that the the fixed point locus is given by \cite{Minahan:2015jta}
\begin{align} 
v^\mu F_{\mu\nu} &=0 \label{eq:locusFv}\, ,\\
\hat{F}^-_{\mu\nu} &=  D_\sigma \tensor{\Phi}{_\mu_\nu^\sigma} \label{eq:locusFhat}\, ,\\
\tilde{F} &=- \tfrac{1}{12} [\Phi_{\mu\nu\la},\tensor{\Phi}{^\mu^\nu_\sigma}](d\k)^{\la\sigma} \, , \label{eq:locusFtilde}\\
v^\sigma \chi_{\sigma \mu\nu \lambda} &=0 \, ,\label{eq:locusChi}\\
 D_\mu \phi_0 &= 0 \label{eq:locusPhi0}\, ,\\
K^m &=-4 \phi_0 (\nu_m \La \eps) \label{eq:locusK}\, ,
\end{align}
where
\begin{equation} \label{chidef}
\chi_{\sigma\mu\nu\lambda} = D_\sigma \Phi_{\mu\nu\lambda} -D_\mu \Phi_{\sigma\nu\lambda}-D_\nu \Phi_{\mu\sigma\lambda}-D_\lambda \Phi_{\mu\nu\sigma}\, ,
\end{equation}
and we have decomposed $F$ as in \eqref{eq:2FormDecomp5}. Equivalently we can rewrite (\ref{bsrt-term}) in terms of differential forms and use the cohomological complex. 

The action restricted to the fixed point locus is \cite{Minahan:2015jta}
\begin{align}
S_{f.p.} = \frac{1}{g^2_7} \int d^7 x \sqrt{-g} \Tr \left(  24 \phi_0\phi_0 +
\tfrac14 F_{\mu\nu} (\iota_R * (F \wedge d\k))^{\mu\nu}  \right)\, .
\end{align}

We can write this in terms of differential forms as
\begin{align}
S_{f.p.} = \frac{1}{g^2_7} \left[ \int 24\, vol_g \Tr(\phi_0^2)  + \frac12 \int \Tr \left( \k \wedge d\k \wedge F \wedge F \right) \right]  \, ,
\end{align}
where $vol_g$ denotes the volume form with respect to the metric $g$. Here we used the identity \eqref{ivstar}.

\subsection{Gauge fixing}
We also need to gauge fix the theory. This is done in the same way as in \cite{Pestun:2007rz}: we introduce the Faddeev-Popov ghosts $c,\bar{c}$ and the Lagrange multiplier $b$ (corresponding to the Lorenz gauge $d^\dagger A = 0$). We also have the zero modes $a_0,\bar{a}_0,b_0,c_0,\bar{c}_0$. The $a$'s and $b$'s are bosonic and the $c$'s fermionic. 

We also introduce a BRST transformation $\db$ and define the combined transformation $Q = \deps + \db$. By arguments similar to those in \cite{Pestun:2007rz} we obtain the following $(Field,Field')$ doublets\footnote{Here the $\tilde{\psi}$ etc, means that we have redefined these fields. However, their exact form are not important for the discussion here.}
\begin{align}
(A,\tilde{\psi}),(\Phi,\tilde{\eta}), (\Upsilon,\tilde{H}), (c,\tilde{\phi_0}), (\bar{c},b), (b_0,c_0), (\bar{a}_0,\bar{c}_0) 
\end{align}
satisfying the canonical $Q$-transformations
\begin{align}
Q Field &= Field'\, , \nn\\
Q Field'&=  (-\L_R + iG_{a_0}) Field \, .
\end{align}

The combined transformation $Q$ thus squares to a Lie derivative and a gauge transformation
\begin{equation}
Q^2 = -\L_R +iG_{a_0}\, .
\end{equation}
This $Q$ will be our odd transformation for the localisation procedure.

In the previous subsection we found the fixed point locus in terms of the transformation $\deps$. The gauge-fixing and setting $Q=\deps+\db$ does not change the locus, apart from that we now identify $\phi_0$ with $a_0$ \cite{Pestun:2007rz}.

\subsection{Localisation of partition function}

In order to calculate the full answer we have to analyse the localisation locus \eqref{eq:locusFv}-\eqref{eq:locusK}. This is a complicated problem and we will make some comments regarding this in the next section. Thus we concentrate on the contribution of flat connections which implies that $A=0$ \footnote{We assume that the Sasaki-Einstein manifold is simply-connected.}. 
Taking $A=0$, $\Phi=0$ and $\phi_0= \sigma =constant $ as an isolated solution to the localisation locus equations \eqref{eq:locusFv}-\eqref{eq:locusK} we obtain the full perturbative partition function from the one-loop approximation:
\begin{equation} \label{partfun1}
\int\limits_g d\sigma e^{- \frac{24}{g^2_7} V_7 \Tr(\sigma^2)} \frac{ \overbrace{\sqrt{det_{\Omega_H^{(2,0)}}(Q^2)det_{\Omega_H^{(0,2)}}(Q^2) det_{\Omega^{0}}(Q^2)}}^{\Upsilon}\overbrace{\sqrt{det_{\Omega^{0}}(Q^2)}}^{c}
\overbrace{\sqrt{det_{\Omega^{0}}(Q^2)}}^{\bar{c}} }{
\underbrace{\sqrt{det_{\Omega^1}(Q^2)}}_{A}
\underbrace{\sqrt{det_{\Omega_H^{(3,0)}}(Q^2)det_{\Omega_H^{(0,3)}}(Q^2)}}_{\Phi}
\underbrace{\sqrt{det_{H^0}(Q^2)}}_{b_0}
\underbrace{\sqrt{det_{H^0}(Q^2)}}_{\bar{a}_0}}\, ,
\end{equation}
where $V_7$ is the volume of the 7-dimensional manifold. Note that the forms are Lie algebra valued, so there are also determinants over the adjoint representation of the Lie algebra which we have not written out in the above expression. Since by now this is a standard calculation, we will be short in our presentation. 

Above we have indicated from which fields the various determinant factors come from. For example, we saw that the fermionic 2-form $\Upsilon$ split into $(2,0)$ and $(0,2)$-forms and a $(1,1)$-form. However, the $(1,1)$ part was proportional to $d\k$, i.e. of the form $fd\k$ for some function $f$, and hence we get a $det_{\Omega^{0}}$-factor corresponding to the $f$.

Firstly, we note that the only harmonic functions on a compact manifold are constants, these are uncharged under $Q^2$, and thus we discard the $\sqrt{det_{H^0}(Q^2)}$ terms. Secondly, if we are interested in the absolute value only, we can ignore phases of the determinants and write $\sqrt{det_{\Omega_H^{(2,0)}}(Q^2)det_{\Omega_H^{(0,2)}}(Q^2)} = det_{\Omega_H^{(0,2)}}(Q^2)$, etc. Finally, decomposing $\Omega^1 = \Omega^0 \kappa \oplus \Omega_H^{(1,0)} \oplus \Omega_H^{(0,1)}$ we can cancel some terms to obtain

\begin{equation} \label{partfun2}
\int\limits_g d\sigma e^{-\frac{24}{g^2_7} V_7 \Tr(\sigma^2)}  \frac{det_{\Omega^{0}}(Q^2) det_{\Omega_H^{(0,2)}}(Q^2)}{det_{\Omega_H^{(0,1)}}(Q^2)det_{\Omega_H^{(0,3)}}(Q^2)}
= \int\limits_g d\sigma e^{-\frac{24}{g^2_7} V_7 \Tr(\sigma^2)}  det_{adj}^{'}\, sdet_{\Omega_H^{(0,\bullet)}}(Q^2)\, ,
\end{equation}
where we wrote out the determinant over the adjoint representation of the Lie algebra explicitly in the last step.
Thus structurally the answer is analogous to that in 5D.

Our aim is now to calculate the superdeterminant $sdet_{\Omega_H^{(0,\bullet)}}(Q^2) = sdet_{\Omega_H^{(0,\bullet)}}(-\L_R + iG_{a_0})$.

As pointed out by Schmude \cite{Schmude:2014lfa}, for Sasaki-Einstein manifolds $X$, this superdeterminant can be obtained by counting holomorphic functions on the metric cone, $C(X)$. Here we give a summary of the argument and refer the reader to \cite{Schmude:2014lfa, Qiu:2016dyj} for more details.

Recall that if $X$ is a Sasaki-Einstein manifold, then the metric cone $C(X)$ is a Calabi-Yau manifold. The Calabi-Yau structure on $C(X)$ gives rise to many nice properties on $X$. For example, we get a contact form $\k$ and its associated Reeb vector field $R$.
The contact form singles out a horizontal space, $X_H$. As described in for example \cite{Sparks:2010sn}, we have a complex structure on $X_H$, and we can define Dolbeault operators. The Dolbeault operator $\bar{\partial}_H: \Omega_H^{(p,q)} \rightarrow \Omega_H^{(p,q+1)}$ gives rise to the Kohn-Rossi cohomology groups, $H_{KR}^{(p,q)}(X)$ \cite{KohnRossi}.

As argued in \cite{Schmude:2014lfa}, $\L_R$ commutes with $\bar{\partial}_H$ and the superdetrminant $sdet_{\Omega_H^{(0,\bullet)}}(-\L_R + iG_{a_0})$ reduces to the same superdeterminant over $H_{KR}^{(0,\bullet)}$. 
Moreover, from the Calabi-Yau structure on $C(X)$ one can obtain a nowhere vanishing horizontal $(3,0)$-form $\Omega$ on $X_H$. This provides a pairing between $(0,0)$-forms and $(0,3)$-forms via $f  \rightarrow \bar{f}\bar{\Omega}$. As argued in \cite{Qiu:2016dyj} this is also a pairing in cohomology, i.e. $H_{KR}^{(0,3)} \iso H_{KR}^{(0,0)}$. However, the $\L_R$ eigenvalues are changed by an overall minus sign and a shift corresponding to the eigenvalue of $\Omega$. 

As in \cite{Qiu:2014oqa} we also have a pairing $H_{KR}^{(0,1)} \iso H_{KR}^{(0,2)}$, but by simply-connectedness the former is zero.
Thus the whole calculation boils down to finding $H_{KR}^{(0,0)}$ (and the shift in charges due to the pairing via $\Omega$). However, $H_{KR}^{(0,0)} \iso H^0(\O_{C(X)})$, so this is the same as counting the holomorphic functions on the cone.

If in addition $X$ is \emph{toric}, this count has a very nice combinatorial description. In section \ref{sec:toric-SE} we discuss this case.

\section{Localisation locus equations}\label{sec:equations}

 In this section we briefly discuss the localisation locus equations  \eqref{eq:locusFv}-\eqref{eq:locusK} and the related issues. 
 Unfortunately  we are not able to say much about these equations, but we make some observations.

\subsection{Contact instantons} \label{sec:contactinst}

If we look at the equations  \eqref{eq:locusFv}-\eqref{eq:locusK} and we set all fields to zero except the connection $A$, we end up with the notion of 7D contact instantons. These can be thought of as a lift of the 4-dimensional instanton equation $*F = \pm F$ to a 7-dimensional contact metric manifold $X$. The arguments here are similar to the 5-dimensional case in \cite{Kallen:2012cs}.

Using the projectors \eqref{eq:HVProjDef} we decompose the field strength into vertical and horizontal components
\begin{equation}
F= F_V+ F_H \, ,
\end{equation}
and by \eqref{ivstar} these are orthogonal with respect to the standard inner product on forms.

One can then decompose the standard Yang-Mills action and obtain the bound
\begin{equation}\label{firstdecomp}
\int\limits_{X} \Tr (F\wedge * F) =  \int\limits_{X} \Tr (F_V \wedge * F_V) + \int\limits_{X} \Tr (F_H\wedge * F_H) \ge \int\limits_{X} \Tr (F_H\wedge * F_H)\, ,
\end{equation}
with equality in the last step if $F_V=0$.

The horizontal part can be decomposed as in \eqref{eq:2FormDecomp5}:
\begin{align}
F_H = \hat{F}^+ + \hat{F}^- +\tfrac{1}{24}\tilde{F} d\k \, ,
\end{align}
and we get
\begin{align}
\int\limits_{X} \Tr (F_H\wedge * F_H) \geq  \int\limits_{X} \Tr (\hat{F}_H^+ \wedge *\hat{F}^+_H) +\int\limits_{X}\Tr (\hat{F}_H^- \wedge *\hat{F}^-_H)\, ,
\end{align}
with equality when $\tilde{F}=0$.

One can further calculate that
\begin{equation}
\int\limits_{X} \Tr (\hat{F}^\pm_H\wedge * \hat{F}^\pm_H) = \frac{1}{2} \int\limits_{X} \Tr (\hat{F}_H\wedge * \hat{F}_H) \pm  \frac{1}{4} \int\limits_{X} \Tr (\kappa \wedge d\k \wedge \hat{F} \wedge \hat{F})\, ,
\end{equation}
and so 
\begin{align}
 \int\limits_{X} \Tr (\hat{F}_H^+ \wedge * \hat{F}^+_H) +\int\limits_{X}\Tr (\hat{F}_H^- \wedge * \hat{F}^-_H)
&\ge \left |  \int\limits_{X} \Tr (\hat{F}_H^+ \wedge *\hat{F}^+_H) - \int\limits_{X}\Tr (\hat{F}_H^- \wedge * \hat{F}^-_H)  \right | \notag \\ 
&= \frac{1}{2} \left | \int\limits_{X}\Tr (\kappa \wedge d\k \wedge \hat{F} \wedge \hat{F} )\right |\, .
\end{align}
In total, we get the following bound for the 7-dimensional Yang-Mills action:
\begin{equation} \label{7dYMbound}
\int\limits_{X} \Tr (F\wedge * F)\ge \frac{1}{2}\left | \int\limits_{X}\Tr (\kappa  \wedge d\k \wedge \hat{F} \wedge  \hat{F})\right |\, .
\end{equation}
Let us discuss this bound, which is somewhat similar to the standard instanton bound in 4D. In 4D this bound is saturated on instantons and making the instanton small does not cost any energy. The instantons can become point-like and exactly these configurations tend to dominate the path integral in supersymmetric gauge theories. Now let us apply similar logic in 7D. In \eqref{7dYMbound} the left hand side is a quasi-topological term that depend only on the contact form $\k$. If we look at when the bound is saturated and we try to `localise' them without any cost in energy then the corresponding configurations should have codimension 4. Thus they should be membrane-like configurations. The configurations with lower codimension have zero Yang-Mills action. For example, the point-like configurations in 7D will have zero Yang-Mills action and we will not be able to invent any topological or quasi-topological terms which will measure them. Therefore it is natural to expect that our 7D Yang-Mills theory has something to do with counting membranes. 

Let us make some final comments about the bound.
The bound \eqref{7dYMbound} is similar to the bound for the five-dimensional case \cite{Kallen:2012cs} and it is saturated when $F_V =0$, $\tilde{F}= 0$ simultaneously with $\hat{F}^-_H = 0$ or $\hat{F}^+_H = 0$. Using the orthogonality of the decomposition \eqref{eq:2FormDecomp} these conditions can be compactly written as the equation
\begin{equation}\label{contactinstanton}
\iota_R * (d\k \wedge F)  = \pm 2 F,
\end{equation}
or equivalently
\begin{equation}\label{contactinstantonV2}
*F = \pm \tfrac12 \k \wedge d\k \wedge F\, .
\end{equation}
Solutions to the corresponding equation in 5D were termed `contact instantons' in \cite{Kallen:2012cs} and were further studied in e.g. \cite{Baraglia}. 
The equation \eqref{contactinstantonV2} with the plus sign is equivalent to the localisation locus equations \eqref{eq:locusFv}-\eqref{eq:locusK} with all fields set to zero except the connection.

We note that the 7D contact instanton equation \eqref{contactinstantonV2} automatically imply the Yang-Mills equation
\begin{equation} \label{eq:YM}
d_A * F =0\, .
\end{equation}
To see this, note that the instanton equation implies that $F\in \hat{\Omega}_H^{2\pm}$ and from the orthogonality of this space to $\check{\Omega}^2_H = \Omega_0 d\k$ one can show that $d\k \wedge d\k \wedge F = 0$. We then get that 
\begin{equation}
d_A * F = \pm \tfrac12 dk \wedge d\k \wedge F = 0\, .
\end{equation}

\subsection{Contact Higgs-Yang-Mills system}

Let us rewrite the fixed point locus \eqref{eq:locusFv}-\eqref{eq:locusK} in a more geometrical form.
The equation \eqref{eq:locusK} fixes the auxiliary field $K$ in terms of the scalar $\phi_0$. The remaining equations can be written as
\begin{align}
\iota_R F & = 0  \, ,\\
\iota_R d_A \Phi &= 0 \, , \\
d_A \phi_0 &=0 \, ,\\
\hat{F}_H^- &= -d^\dagger_A \Phi\, ,\\
\check{F}_H \wedge d\k \wedge d\k &= 4 [\Phi^{-}, \Phi^+] \, .
\end{align}

Here we have decomposed the field strength as in \eqref{eq:2FormDecomp}, $F = F_V + \hat{F}_H^{+}+ \hat{F}_H^{-} + \check{F}_H$, where $\hat{F}_H^-$ decomposes as a $(2,0)$ and a $(0,2)$ form and $\check{F}_H$ is a $(1,1)$-form proportional to $d\k$.
The horizontal 3-form $\Phi$ is decomposed as in \eqref{eq:3FormDecomp}, $\Phi = \Phi^+ +\Phi^-$, where $\Phi^+$ is a $(0,3)$-form and $\Phi^-$ a $(3,0)$-form.
The first two equations say that $F$ and $d_A \Phi$ are horizontal, and the remaining equations correspond to the 6D Hermitian Higgs-Yang-Millls system studied in e.g. \cite{Iqbal2003}.
These equations are a natural lift of the 6D Hermitian Higgs-Yang-Millls system to a contact 7D manifold. Indeed they are well-defined for any 7D K-contact manifold. One can further 
rewrite them using tricks from the previous subsection and write energy bounds which involve the $\Phi$ field. Hopefully these equations are mathematically interesting.

\section{Localisation for  toric Sasaki-Einstein manifolds}\label{sec:toric-SE}

In section \ref{sec:localization} we obtained the perturbative part of the partition function in terms of the superdeterminant $sdet_{\Omega_H^{(0,\bullet)}}(-\L_R + iG_{a_0})$. For Sasaki-Einstein manifolds $X$ we saw that the calculation of this superdeterminant reduced to counting holomorphic functions on the metric cone $C(X)$. For toric Sasaki-Einstein manifolds this count can be done explicitly, and this is the topic of this section.

\subsection{Toric Sasaki-Einstein manifolds}

We begin by summarising some properties of toric Sasaki-Einstein manifolds. A good introduction to this subject can be found in e.g. \cite{Sparks:2010sn, dasilva}.

We say that a $7$-dimensional Sasaki-Einstein manifold $X$ is \emph{toric} if its metric cone $C(X)$ admits an effective Hamiltonian action of the $4$-torus $\T^4$. We also require that the Reeb vector field $R$ lies in the Lie algebra of the torus action, i.e. that the Reeb vector field is a linear combination of the four $U(1)$ actions of the torus.
Associated to the torus action we have a moment map $\mu: C(X) \rightarrow \R^4$ whose image forms a convex rational polyhedral cone, $C_\mu \subset \R^4$, which we will refer to as the \emph{moment map cone}. We can represent $C_\mu$ as
\begin{equation} \label{momentmapcone}
C_\mu = \{ \vec{x} \in \R^4 | \vec{x} \cdot \vec{u}_i \geq 0\}\, ,
\end{equation}
where $\vec{u}_i$, $i=1,\dots,d$ are the inward pointing normals of the hyperplanes bounding $C_\mu$.
The rationality property allows us to pick $\vec{u}_i \in \Z^4$ and we can take them to be primitive, meaning that all components in a given $\vec{u}_i$ are relatively prime. The cone is also \emph{good} in the sense of Lerman \cite{Lerman}, see Appendix \ref{AppSine}, which in particular means that the manifold itself can be reconstructed from the cone. 

Much of the information about the manifold can be read off directly from the moment map cone. For us, the interest lies in the holomorphic functions on $C(X)$ and their eigenvalues under $\L_R$. The holomorphic functions correspond to integer lattice points in the moment map cone, and their charges under the four $U(1)$'s of the torus action are given by the coordinates of those lattice points. Since the Reeb vector field is a linear combination of the four $U(1)$'s one can also read off the $\L_R$ eigenvalues. If we represent the Reeb by a 4-vector $\vec{R}$, then for a holomorphic function corresponding to the lattice point $\vec{n} \in C_\mu \cap \Z^4$ its eigenvalue is simply given by $\vec{n} \cdot \vec{R}$. 

For our superdeterminant calculation we also need to find the shift due to the holomorphic volume form $\Omega$ that pairs (0,0)-forms and (0,3)-forms. The Sasaki-Einstein condition on $X$ means that there exists a primitive vector $\vec{\xi}$ such that $\vec{\xi} \cdot \vec{u}_i = 1$ for all $\vec{u}_i$, $i=1,\dots,d$. This is known as the 1-Gorenstein condition \cite{Sparks:2010sn}. The $\L_R$ eigenvalue of $\Omega$ is then given by $\vec{\xi} \cdot \vec{R}$.

Ignoring the Lie algebra part for the moment, we can write our superdeterminant in terms of the moment map cone:
\begin{align}\label{superdeterminant}
sdet_{\Omega_H^{(0,\bullet)}}(-\L_R + x) &= \frac{\prod \limits_{\vec{n}\in C_\mu(X) \cap \Z^4 \setminus \{\vec{0}\}}\left( \vec{n} \cdot \vec{R} + x \right) } {\prod \limits_{\vec{n}\in C_\mu(X) \cap \Z^4 \setminus \{\vec{0}\}}\left(- \vec{n} \cdot \vec{R} - \vec{\xi} \cdot \vec{R} + x \right)}\, .
\end{align}
Note that we removed the origin since $\vec{n}=\vec{0}$ corresponds to a constant function with zero charge.

These infinite products are of course divergent but can be $\zeta$-function regularised.

Using the 1-Gorenstein condition, $\vec{\xi} \cdot \vec{u}_i = 1$, we can rewrite the denominator to obtain
\begin{align}
sdet_{\Omega_H^{(0,\bullet)}}(-\L_R + x) &= \frac{\prod \limits_{\vec{n}\in C_\mu(X) \cap \Z^4 \setminus \{\vec{0}\}}\left( \vec{n} \cdot \vec{R} + x \right) } {\prod \limits_{\vec{n}\in C^\circ_\mu(X) \cap \Z^4}\left( \vec{n} \cdot \vec{R} - x \right)}\, ,
\end{align}
where $C^\circ_\mu(X)$ denotes the interior of the moment map cone. Here we also absorbed the minus signs in the denominator since they do not matter once you regularise.

Before proceeding to some examples, let us re-instate the Lie algebra components and state the final answer. Let $\beta$ denote the non-zero roots of the Lie algebra $g$ and let $t$ denote the Cartan subalgebra. Using the Weyl integration formula we can rewrite \eqref{partfun2}, up to some overall factors, as
\begin{align}
Z^{\text{pert}}_X&=\int\limits_t d \sigma\,  e^{-\frac{24}{g^2_7} V_7 \Tr(\sigma^2)} \prod_\beta   i \bra \sigma, \beta \ket \frac{\prod \limits_{\vec{n}\in C_\mu(X) \cap \Z^4 \setminus \{\vec{0}\}}\left( \vec{n} \cdot \vec{R} +  i \bra \sigma, \beta \ket \right) } {\prod \limits_{\vec{n}\in C^\circ_\mu(X) \cap \Z^4}\left( \vec{n} \cdot \vec{R} -  i \bra \sigma, \beta \ket \right)} \nonumber\\
&= \int\limits_t d \sigma \,  e^{-\frac{24}{g^2_7} V_7 \Tr(\sigma^2)} \prod_\beta S_4^{C_\mu(X)}( i \bra \sigma, \beta \ket |\vec{R})\, , \label{pertparfun}
\end{align}
where
\begin{equation}
S_4^{C_\mu(X)} (x|\vec{R}) = \frac{\prod \limits_{\vec{n}\in C_\mu(X) \cap \Z^4}\left( \vec{n} \cdot \vec{R} +  x \right) } {\prod \limits_{\vec{n}\in C^\circ_\mu(X) \cap \Z^4}\left( \vec{n} \cdot \vec{R} -  x \right)} \, .
\end{equation}
This function $S_4^{C_\mu(X)}$ is known as the \emph{generalised quadruple sine function} associated to the cone $C_\mu(X)$, see Appendix \ref{AppSine}.

\subsubsection{The 7-sphere} \label{7-Sphere}

As the simplest example of the above, let us consider the 7-sphere, i.e. $X=S^7$.
In this case $C(X)$ is simply $\C^4$ whose coordinates we will denote by $(z_1,z_2,z_3,z_4)$.
Let $T$ be the torus action whose four $U(1)$'s rotate four $z_i$'s separately, i.e. 
\begin{align}
T:(z_1,z_2,z_3,z_4) \mapsto (e^{i\phi_1}z_1,e^{i\phi_2}z_2,e^{i\phi_3}z_3,e^{i\phi_4}z_4)\, .
\end{align}
The corresponding moment map is 
\begin{align}
\mu: (z_1,z_2,z_3,z_4) \mapsto (|z_1|^2,|z_2|^2,|z_3|^2,|z_4|^2)
\end{align}
 and the moment map cone is given by the positive orthant of $\R^4$, i.e. 
\begin{align}
 C_\mu = \R_{\geq 0}^4 = \{ (x_1,x_2,x_3,x_4)\in \R^4 | x_i \geq 0 \}\, .
\end{align} 
The integer lattice points inside the moment map cone are thus $\Z_{\geq 0}^4$ and they are in one-to-one correspondence with the holomorphic functions on $\C^4$ via $(n_1,n_2,n_3,n_4) \leftrightarrow z_1^{n_1}z_2^{n_2}z_3^{n_3}z_4^{n_4}$. 

The inward pointing normals $\{\vec{u}_i\}$ of $C_\mu = \R_{\geq 0}^4$ can be taken to be the standard basis of $\R^4$ and the dual of the moment map cone is thus also $\R_{\geq 0}^4$.
If we denote by $e_i$ the vector fields associated to each $U(1)$ action, we can write the Reeb vector as $R = \sum_{i=1}^4  \w_i e_i$ and associate it to the vector $\vec{R} = (\w_1,\w_2,\w_3,\w_4)$ in $\R_{>0}^4$. For a general $\vec{R}$ we get a squashed sphere and setting all $\w_i=1$ results in the round sphere. 

Each lattice point $\vec{n} = (n_1,n_2,n_3,n_4) \in \Z_{\geq 0}^4$ thus contributes with a factor of $\vec{n} \cdot \vec{R}=n_1\w_1+n_2\w_2+n_3\w_3+n_4\w_4$ to the superdeterminant.

For $C(X)=\C^4$ the holomorphic volume form $\Omega$ is simply given by $\Omega = dz_1 \wedge dz_2 \wedge dz_3 \wedge dz_4$ and its eigenvalue under $\L_R$ is $\w_1+\w_2+\w_3+\w_4$. In terms of our moment map cone, this eigenvalue is obtained in the following way:
The vector $\vec{\xi}$ satisfying $\vec{\xi} \cdot \vec{u_i}=1$ is $\vec{\xi} = (1,1,1,1)$ and taking the dot product with $\vec{R}$ we get $\vec{\xi} \cdot \vec{R} = \w_1+\w_2+\w_3+\w_4$.

In the case of $X=S^7$ we can thus write the pertubative part of the partition function \eqref{pertparfun} as \cite{Minahan:2015jta}
\begin{align}
Z^{\text{pert}}_{S^7}=\int\limits_t d \sigma e^{-\frac{24}{g^2_7} V_7 \Tr(\sigma^2)} \prod_\beta S_4( i \bra \sigma, \beta \ket | \vec{R})\, ,
\end{align}
where
\begin{equation}
S_4 (x|\vec{R}) = \frac{\prod\limits_{n_1,n_2,n_3,n_4 \geq 0}\left( n_1\w_1+n_2\w_2+n_3\w_3+n_4\w_4 + x \right) } {\prod\limits_{n_1,n_2,n_3,n_4 \geq 1}\left( n_1\w_1+n_2\w_2+n_3\w_3+n_4\w_4 - x \right)} 
\end{equation}
is the ordinary quadruple sine function, see Appendix \ref{AppSine}.

For the round sphere where all $\w_i=1$ \footnote{This Reeb vector field corresponds to  the Sasaki-Einstein geometry.} the quadruple sine reduces to
\begin{equation}
S_4 (x) = \frac{\prod\limits_{n_1,n_2,n_3,n_4 \geq 0}\left( n_1 + n_2 + n_3 +n_4 + x \right) } {\prod\limits_{n_1,n_2,n_3,n_4 \geq 1}\left( n_1 + n_2 + n_3 + n_4 - x \right)}
= x \prod_{t=1}^\infty \frac{(t+x)^{\frac{1}{6}t^3+t^2+\frac{11}{6}t +1}}{(t-x)^{\frac{1}{6}t^3-t^2+\frac{11}{6}t -1}} \, ,
\end{equation}
which can be interpreted in terms of the Hopf fibration of the round sphere, see \cite{Minahan:2015jta}. For a discussion of the analytical continuation of this answer to other dimensions, see \cite{Minahan:2015any, Minahan:2017wkz}.

\subsubsection{Examples of toric Sasaki-Einstein manifolds}

Let us provide a few other explicit examples for toric Sasaki-Einstein manifolds. 
There are plenty of examples of toric  Sasaki-Einstein manifolds in the literature. In \cite{Gauntlett:2004yd} it was shown that from any positive curvature K\"ahler-Einstein manifold one can construct an infinite class of Sasaki-Einstein manifolds in three dimensions higher. In \cite{Martelli:2008rt} this construction was applied to $\C P^2$ and $\C P^1 \times \C P^1$  to obtain two infinite families of seven-dimensional Sasaki-Einstein manifolds denoted $Y^{p,k}(\C P^2)$ and $Y^{p,k}(\C P^1 \times \C P^1)$ respectively \footnote{Our discussion here will be brief and only concern facts necessary for our purpose. For a more thorough description we refer the reader to \cite{Martelli:2008rt}.}. These were also given a toric description in terms of K\"ahler quotients, and the corresponding moment map cones were found. We will write the answer for a generic toric Reeb.

For $Y^{p,k}(\C P^2)$ the moment map cone is described as in \eqref{momentmapcone} by the inward-pointing normal vectors \cite{Martelli:2008rt}
\begin{align}
& & \vec{u}_1 = (0,0,-1,0)\, , & & \vec{u}_2 = (0,0,-1,-p)\, , & & \vec{u}_3 = (-1,0,-1,0)\, , \nn\\
& & \vec{u}_4 = (0,-1,-1,0)\, , & & \vec{u}_5 = (1,1,-1,-k)\, .
\end{align}
The vector $\vec{\xi}$ satisfying $\vec{\xi} \cdot \vec{u_i}=1$ is $\vec{\xi}=(0,0,-1,0)$. One can find the Reeb vector as a linear combination of the four $U(1)$ actions. For the squashed case the generic Reeb can be written as $R = \sum_{i=1}^4  \w_i e_i$ where $e_i$ are the basis vectors associated to each $U(1)$, and $\vec{R} = (\w_1,\w_2,\w_3,\w_4)$ satisfies the dual cone condition $\vec{R} = \sum \limits_i \lambda_i \vec{u}_i$ for some $\lambda_i>0$.

One also has to find the vectors $\vec{n}=(n_1,n_2,n_3,n_4)$ corresponding to the integer lattice points inside the moment map cone. This is done by solving $\vec{n} \cdot \vec{u}_i \geq 0 $ and the quadruple sine in the perturbative part of the partition function \eqref{pertparfun} becomes
\begin{align}\label{scp2}
S_4^{C_\mu (Y^{p,k}(\C P^2))}(x|\vec{R})= \frac{ \displaystyle \smashoperator[l]{\prod\limits_{n_3 =-\infty}^{0}} \prod\limits_{n_4= -\infty}^{-\delta} \prod\limits_{n_2=2n_3 +kn_4}^{-n_3}  \prod\limits_{n_1 = n_3+k n_4-n_2 }^{-n_3} (n_1 \omega_1 + n_2 \omega_2 + n_3 \omega_3 + n_4 \omega_4 +x)}{ \displaystyle \smashoperator[l]{\prod\limits_{n_3 =-\infty}^{0}} \prod_{n_4= -\infty}^{-\delta} \prod\limits_{n_2=2n_3 + kn_4}^{-n_3}  \smashoperator[r]{\prod\limits_{n_1 = n_3+k n_4 -n_2}^{-n_3}} (n_1 \omega_1 + n_2 \omega_2 + (n_3+1) \omega_3 + n_4 \omega_4 -x)}\, ,
\end{align}
where $\delta= \lceil \frac{n_3}{p} \rceil$. For a given $p$, $k$ lies in the range $\frac{3p}{2}\leq k \leq 3 p$ \cite{Martelli:2008rt}, and fixing specific values one can simplify the above expression further. Note that the answer is symmetric under the interchange of $\omega_1$ and $\omega_2$ although this symmetry is obscured by the way we chose to parametrise the bounds in \eqref{scp2}.

For $Y^{p,k}(\C P^1 \times \C P^1)$ the inward pointing normals are \cite{Martelli:2008rt}
\begin{align}
& & \vec{u}_1 = (0,0,-1,0)\, & & \vec{u}_2 = (0,0,-1,-p)\, , & & \vec{u}_3 = (1,0,-1,0)\, , \nn\\
& & \vec{u}_4 = (-1,0,-1,-k)\, , & & \vec{u}_5 = (0,1,-1,0)\, , & & \vec{u}_6 = (0,-1,-1,-k)\, .
\end{align}
Here we also have $\vec{\xi}=(0,0,-1,0)$. Writing the general Reeb vector as before and parametrising the integer lattice points in the moment map cone, we get
\begin{align} \label{eq:S4Ypkcp1}
S_4^{C_\mu (Y^{p,k}(\C P^1 \times \C P^1))}(x|\vec{R})= \frac{\smashoperator[l]{\prod\limits_{n_3 =-\infty}^{0}} \prod\limits_{n_4= -\infty}^{-\delta} \prod\limits_{n_1 =  n_3}^{-n_3-k n_4} \prod\limits_{n_2 =  n_3}^{-n_3-k n_4}  (n_1 \omega_1 + n_2 \omega_2 + n_3 \omega_3 + n_4 \omega_4 +x)}{\smashoperator[l]{\prod\limits_{n_3 =-\infty}^{0}} \prod\limits_{n_4= -\infty}^{-\delta} \prod\limits_{n_1 =  n_3}^{-n_3-k n_4} \prod\limits_{n_2 =  n_3}^{-n_3-k n_4} (n_1 \omega_1 + n_2 \omega_2 + (n_3+1) \omega_3 + n_4 \omega_4 -x)}\, ,
\end{align}
where again $\delta= \lceil \frac{n_3}{p} \rceil$. For specific $p$, $k$ lies in the range $p\leq k \leq 2 p$ \cite{Martelli:2008rt}, and for specific values on can write \eqref{eq:S4Ypkcp1} in a nicer form. As an example, for $p=1$ we can rewrite the answer in terms of ordinary quadruple sines
\begin{align}\label{s4cp1cp1}
S_4&^{C_\mu (Y^{1,k}(\C P^1 \times \C P^1))}(x|\vec{R})= 
\frac{S_4(x|\w_1,\w_2,-\w_1-\w_2-\w_3+\w_4, -\w_4)}{S_4(x+\w_2| \w_1, \w_2,-\w_1-\w_3+\w_4 +(1-k)\w_2, k\w_2-\w_4)} \nonumber \\
&\quad \times
\frac{S_4(x+\w_1+\w_2|\w_1,\w_2,-\w_3+\w_4+(1-k)\w_1 +(1-k)\w_2, k\w_1+k\w_2-\w_4)}{S_4(x+\w_1|\w_1, \w_2,-\w_2-\w_3+\w_4+(1-k)\w1, k\w_1-\w_4)}\, .
\end{align}

The special case $Y^{1,1}(\C P^1 \times \C P^1)$ is known in the literature as $Q^{1,1,1}$ and is a $U(1)$-bundle over $\C P^1 \times\C P^1 \times\C P^1$ (see e.g. \cite{Fabbri1999} for a description). Setting $k=1$ in
\eqref{s4cp1cp1} we have

\begin{align}
S_4^{C_\mu (Q^{1,1,1})}(x|\vec{R})&= 
\frac{S_4(x|\w_1,\w_2,-\w_1-\w_2-\w_3+\w_4, -\w_4)}{S_4(x+\w_2|\w_1, \w_2,-\w_1-\w_3+\w_4, \w_2-\w_4)} \nn \\
&\quad \times
\frac{S_4(x+\w_1+\w_2|\w_1,\w_2,-\w_3+\w_4, \w_1+\w_2-\w_4)}{S_4(x+\w_1|\w_1, \w_2,-\w_2-\w_3+\w_4, \w_1-\w_4)}\, .
\end{align}

\subsection{Factorisation}\label{Factor}

In the above discussions we allowed for general Reebs $\vec{R} = (\w_1,\w_2,\w_3,\w_4)$ within the dual moment map cone. However, the generalised quadruple sine functions are also defined for complex $\vec{R}$ as long as the real part can be rotated to lie in the dual cone \cite{Winding:2016wpw}. Thinking of our quadruple sines as complex functions we can use factorisation results for these special functions and apply them to our perturbative partition function. Having factorised the perturbative part of the partition function we recognise the pieces as perturbative Nekrasov partition functions. It is then natural to conjecture that the full partition function obeys the same factorisation and can be written as a product of full Nekrasov partion functions. This argument is similar to the 5D case discussed in \cite{Qiu:2014oqa} (see also the earlier works \cite{Kim:2012ava, Lockhart:2012vp, Kim:2012qf}).  

We will start by defining some additional notions. We refer the reader to \cite{Winding:2016wpw} for a thorough treatment of generalised quadruple sine functions and their factorisations, some of which are also summarised in Appendix \ref{AppSine}.

Let $x=e^{ 2\pi i z}$, $q_j = e^{2\pi i \t_j}$ where $z, \t_j \in \C$ with $\rm Im (\t_j) \neq 0$, $j=0, \dots, r$. We use the following compact notation
\begin{align}
\qvec=& (q_0, \ldots , q_r)\, ,\\
\qvec^-(j)=& (q_0 , \ldots , q_{j-1},q_{j+1},\ldots ,q_r)\, ,\\
\qvec[j]=& (q_0 , \ldots ,q^{-1}_j,\ldots ,q_r)\, ,\\
\qvec^{-1}=& (q_0^{-1}, \ldots , q_r^{-1})\, .
\end{align}
Assume that $\rm Im (\t_j) < 0$ for $j=0,\dots,k-1$ and $\rm Im (\t_j)>0$ for $j = k, \dots, r$. The $q$-shifted factorial is then defined to be \cite{Narukawa}
\begin{equation}
(x|\qvec)_\infty = \prod_{j_0,\ldots,j_r =0}^\infty \left(1- xq_0 ^{-j_0 -1} \ldots q_{k-1}^{-j_{k-1}-1}q_k^{j_k} \ldots q_r^{j_r}\right)^{(-1) ^k}\, .
\end{equation}
By defining it to also be invariant under re-orderings of $\qvec$, the above expression defines the $q$-shifted factorial for any choice of $\t_j$ with non-zero imaginary part.
The $q$-shifted factorial is a meromorphic function of $z$ and it satisfies some interesting identities like \cite{Narukawa}
\begin{equation}
(x|\qvec)_\infty = \frac{1}{(q_j^{-1} x | \qvec[j])_\infty}\, ,\qquad (q_jx|\qvec)_\infty = \frac{(x|\qvec)_\infty}{( x | \qvec^-(j))_\infty}\,.
\end{equation}

One can use these functions to rewrite the generalised quadriple sine in \eqref{pertparfun}. Following Theorem 4.4 in \cite{Winding:2016wpw} one can factorise a generalised quadruple sine $S^C _4 (x|\vec{\w})$ associated to a good cone\footnote{See Appendix \ref{AppSine} for the definition of a good cone. Recall that the moment map cones associated to our toric Sasaki-Einstein manifolds will have this property.} $C$ into the following product:
\begin{align}
\label{factor}
S_4^C (x|\vec{\w}) =& e^{\frac{\pi i }{4!} B^C _{4,4}(x|\vec{\w})}\prod_{k \in \Delta_1^C} (z_k|\qvec_k)_\infty \nn \\
=& e^{-\frac{\pi i }{4!} B^C _{4,4}(x|\vec{\w})}\prod_{k \in \Delta_1^C} (z_k^{-1}|\qvec^{-1}_k)_\infty \, . 
\end{align}
Let us explain the notation used above. On the left-hand side $x\in \C$ and $\vec{\w}=(\w_1,\w_2,\w_3,\w_4) \in \C^4$. On the right-hand side $B^C_{4,4}$ is a generalised Bernoulli polynomial \cite{Winding:2016wpw}, a fourth-order polynomial in $x$ whose exact form we will omit here. The $\Delta_1^C$ denotes the generators of the cone i.e. the one-dimensional faces or rays of the cone. To each ray $k$ of the cone one can associate an $SL_4(\mathbb{Z})$-matrix $\tilde{K}_k$ (see Appendix \ref{AppSine}) and the above factorisation requires that ${\rm Im }\frac{(\tilde{K}_k \vec{\w})_j}{(\tilde{K}_k \vec{\w})_1}\neq 0$, for $j=2,3,4$. Finally, the terms inside the $q$-shifted factorials are defined as
\begin{equation}\label{qfactorialvariables}
z_k = e^{2\pi i \frac{x}{(\tilde{K}_k \vec{\w})_1}}\quad \text{and} \quad \qvec_k = \left( e^{2\pi i \frac{(\tilde{K}_k \vec{\w})_2}{(\tilde{K}_k \vec{\w})_1}},  e^{2\pi i \frac{(\tilde{K}_k \vec{\w})_3}{(\tilde{K}_k \vec{\w})_1}},  e^{2\pi i \frac{(\tilde{K}_k \vec{\w})_4}{(\tilde{K}_k \vec{\w})_1}}\right)\,.
\end{equation}

The factorisation of the generalised quadruple sine is an important property and it has a nice interpretation in terms of partition functions. To illustrate this, let us first return to the squashed sphere of section \ref{7-Sphere}.

For $S^7$ the moment map cone is given by $\R^4_{\geq 0}$ which has four rays corresponding to the four standard basis vectors in $\mathbb{R}^4$. For this case the matrix $\tilde{K}_1=\rm Id_4$ and the other $\tilde{K}_k$ are obtained by permuting the rows of the identity matrix. When the cone is 
$\R^4_{\geq 0}$ the generalised quadruple sines and Bernoulli polynomials reduce to their ordinary versions and one can easily see that the above result \eqref{factor} reduces to the factorisation of the normal quadruple sine that was described in \cite{Minahan:2015jta}:
\begin{align} \label{factorS7}
S_4 (x|\vec{\w}) =  e^{\frac{\pi i }{4!} B_{4,4}(x|\vec{\w})}\prod_{k =1}^4 (z_k|\qvec_k)_\infty \, ,
\end{align}
where $z_k = e^{2\pi i \frac{x}{\w_k}}$ and $\qvec_k = \left( e^{2\pi i \frac{\w_1}{\w_k}},\dots ,e^{2\pi i \frac{\w_{k-1}}{\w_k}},e^{2\pi i \frac{\w_{k+1}}{\w_k}},\dots,e^{2\pi i \frac{\w_4}{\w_k}}\right)$.

Notice that each of the pieces in the factorisation correspond to a perturbative Nekrasov partition function, $Z^{\text{pert}}_{S^1 \times \C^3}$, on $S^1 \times_\epsilon \C^3$ \cite{Nekrasov:2008kza}. This corresponds to the decomposition of $S^7$ into four pieces around the closed Reeb orbits for the toric Reeb. The neighbourhoods around these orbits have local geometry $S^1 \times_{\epsilon} \C ^3$, where we have imposed a twisted periodic boundary condition along the $S^1$ using three equivariant parameters $\epsilon_i, \, i=1,2,3$ corresponding to the three $U(1)$'s acting on each $\C$. According to \eqref{factorS7} one can write the perturbative part of the partition function as 
\begin{equation}\label{perturbatives1}
Z^{\text{pert}}_{S^7} = \prod_{i=1}^4 Z^{\text{pert}}_{S^1 \times \C ^3} (\beta_i, \epsilon_{1i}, \epsilon_{2i}\, , \epsilon_{3i}),
\end{equation}
where $\beta_i$ is the radius of the corresponding $S^1$. Thus it is natural to say that the full partition function also factorises in the same way, with the building blocks being the full Nekrasov partition function on $S^1 \times \C^3$:
\begin{equation}
Z^{\text{full}}_{S^7} = \prod_{i=1}^4 Z^{\text{full}}_{S^1 \times \C ^3} (\beta_i, \epsilon_{1i}, \epsilon_{2i}\, , \epsilon_{3i})~,\label{full-PF-S7}
\end{equation}
 where the full partition function for the 7D theory on $S^1 \times \C ^3$ is defined in terms 3D-partitions, see \cite{Nekrasov:2008kza}. 

Applying the same logic to the factorisation \eqref{factor} one can conjecture that the full partition function for any 7D toric Sasaki-Einstein manifold $X$ will be  
\begin{equation}
Z^{\text{full}}_{X} = \prod_{k \in \Delta_1^C} Z^{\text{full}}_{S^1 \times \C ^3} (\beta_k, \epsilon_{1k}, \epsilon_{2k}\, , \epsilon_{3k})\, ,\label{full-PF}
\end{equation}
i.e. that it can be written in terms of Nekrasov partition functions on $S^1 \times_\epsilon \C^3$, with one factor for every generating ray of the moment map cone $C$ of $X$, and the parameters can be read off from \eqref{qfactorialvariables}.

These results are very similar to the ones for 5D theories, see for example \cite{Kallen:2012va}, and it is believed that the 5D partition function localises on closed Reeb orbits. For 5D theory the appearance of Reeb orbits as localisation locus is very natural, however in 7D it may appear not to be the full answer. For example, the configurations with support on closed Reeb orbits will have zero Yang-Mills action. The only term which is not zero will be the following quasi-topological term
\begin{equation}\label{quasi-top}
\int\limits_{X} \Tr \left( \k \wedge F \wedge F \wedge F \right)\,. 
\end{equation} 
Thus the conjectures \eqref{full-PF-S7} and \eqref{full-PF} may appear to be somewhat naive and there can be more non-trivial contributions to the full partition function. However at the moment we have no possibility to probe them.

\section{3-Sasaki manifolds}\label{sec:3-Sasaki}

So far we have focused on aspects related to the Sasaki-Einstein structure of our manifolds. However, in dimension 7 some of these manifolds may also be 3-Sasakian. We believe that the additional structures associated to the manifold being 3-Sasakian should manifest itself at the level of partition functions, and we make some conjectures about this at the end of this section.

As mentioned in section \ref{7dkillingspinors}, a 3-Sasakian manifold\footnote{For a thorough introduction to 3-Sasakian geometry we refer the reader to \cite{Boyer:1998sf}.} can be defined as a manifold whose metric cone is hyperk\"ahler. Thus the cone is equipped with three complex structures $I,J, K$ and a metric compatible with all three complex structures. The base of the cone, i.e. the 3-Sasakian manifold, is equipped with three contact forms $\k_a$, $a=1,2,3$, and three Reeb vectors $R_a$ that satisfy the following relations
\begin{align}
[R_a, R_b]&= \epsilon_{abc}R_c\, ,\\
\iota_{R_a}\k_b &= \delta_{ab}\, .
\end{align}
There is also a metric compatible with all three contact forms. For any $\xi \in S^2$ we can define  a contact form $\k = \xi^a \k_a$ and a corresponding Reeb $R=\xi^a R_a$. Thus a 3-Sasakian structure has an $S^2$ worth of contact structures. A 3-Sasaki manifold admits a free action of $SU(2)$ and it is thus a principal $SU(2)$-bundle. 

We will discuss observables for 3-Sasakian manifolds. We will also look at the perturbative answer for the 3-Sasakian case and try to identify $SU(2)$ structures there. We focus our attention to the sphere $S^7$ but we believe that our observations hold for any 3-Sasaki manifold. 

\subsection{7D Observables}

Besides the Yang-Mills action we can construct other actions invariant under supersymmetry. In particular there are interesting cohomological observables which we will discuss briefly. 

Let $X$ be a 7D contact manifold with contact form $\k$ and Reeb vector $R$.
Consider the part of the cohomological complex given by 
\begin{align}
\delta F &= d_A \Psi\, ,\nn\\
\delta \Psi &= \iota_R F+ d_A \sigma\, ,\label{variationnormal}\\ 
\delta \sigma &= \iota_R \Psi\, ,\nn
\end{align}
coming from \eqref{a-variation}-\eqref{s-variation}. Here we have redefined $\sigma \rightarrow -i\sigma$ and reversed the contact structure for notational convenience.

We can rewrite these transformations as
\begin{align}
\delta ~ \Tr (\sigma  + \Psi + F)^k   = (d+\iota_R) ~{\rm Tr} (\sigma + \Psi + F)^k\,.
\end{align} 

We can then introduce the supersymmetrisation of different Chern-Simons-like terms
\begin{align}
S_{7,k}= \int\limits_{X} \left (\Tr  (\sigma + \Psi + F)^k \wedge \Omega \right )\, ,
\end{align}
where $\Omega$ should satisfy certain cohomological conditions in order for $\delta S_{7,k}=0$. 
In 7D we will be allowed to have $k=1$ (FI term), $k=2,3$ and $k=4$ (7D proper Chern-Simons term which should be treated separately within this formalism). If we include these observables into the theory, then we will induce new classical terms, for example linear in $\sigma$ for $k=1$, quadratic in $\sigma$ for $k=2$ (this is the same as the supersymmetric Yang-Mills action), cubic in $\sigma$ for $k=3$ and quartic in $\sigma$ for the supersymmetric 7D Chern-Simons term.  

Let us analyse the conditions for $\Omega$ in some detail. For example, we can choose
\begin{align}
S_{7,2}= \int\limits_{X} \left (\Tr  (\sigma + \Psi + F)^2 \wedge (\Omega_3 + \Omega_5 + \Omega_7) \right )\, ,
\end{align}
and if 
\begin{align}
(d+\iota_R) (\Omega_3 + \Omega_5 + \Omega_7) = \text{any two-form}\, ,
\end{align}
then this is invariant under $\delta$. 
For example, we can take 
\begin{align}
\Omega_3 + \Omega_5 + \Omega_7 = \kappa d\kappa - \kappa (d\kappa)^2 + \kappa (d\kappa)^3\, .
\end{align}
We can also introduce another observable:
\begin{align}\label{observable73}
S_{7,3}= \int\limits_{X} \left (\Tr  (\sigma + \Psi + F)^3 \wedge (\Omega_1+ \Omega_3 + \Omega_5 + \Omega_7) \right )
\end{align}
and $\delta S_{7,3}= 0$ if 
\begin{align}
(d+\iota_R) (\Omega_1+ \Omega_3 + \Omega_5 + \Omega_7) = \text{any function}\,.
\end{align}
An example is given by 
\begin{align}
\Omega_1 + \Omega_3 + \Omega_5 + \Omega_7 = \kappa - \kappa d\kappa + \kappa (d\kappa)^2 - \kappa (d\kappa)^3\, .
\end{align}
Note that the gauge theory part of the above observable \eqref{observable73} is the quasi-topological term \eqref{quasi-top} that gives a non-zero contribution in the partition function. In addition to the observabels above, one can also consider other Chern-Simons-like observables, see \cite{Kallen:2012cs}.

\subsubsection{3-Sasaki manifolds}

Let us generalise the above observables to the case of a 7D 3-Sasaki manifold $X$. We then have three contact forms $\k_a$ and their corresponding Reeb vectors $R_a$, $a=1,2,3$.

Consider the Cartan model of equivariant cohomology with the differential
\begin{align}
d_{\rm eq} = d + \xi^a \iota_{R_a}\, ,
\end{align}
acting on forms which are also polynomials in $\xi$ (which has degree $2$). 
Now the transformations can be defined as follows
\begin{align}
& \delta A = \Psi\, \nn,\\
&\delta \Psi = \xi^a \iota_{R_a} F+ d_A \sigma\, ,\\
& \delta \sigma = \xi^ a\iota_{R_a} \Psi \, .\nn
\end{align}
Notice that by restricting $\xi \in S^2$ one can retrieve the above variation \eqref{variationnormal}, but this is not our aim here and we will allow $\xi^a$ to be arbitrary parameters. Our aim is to make an asantz about supersymmetry transformations that manifest all three contact structures. We can rewrite the transformations as follows
\begin{align}
\delta (\sigma + \Psi + F) = (d_A +  \xi^a \iota_{R_a})  (\sigma + \Psi + F)\, ,
\end{align}
and furthermore
\begin{align}
\delta ~\Tr (\sigma + \Psi + F)^k   = (d+ \xi^a \iota_{R_a} ) ~\Tr (\sigma+ \Psi + F)^k\, . 
\end{align}
We can introduce the observable 
\begin{align}
S_{7,2}= \int\limits_{X} \left (\Tr  (\sigma + \Psi + F)^2 \wedge (\Omega_{3\, ab} \xi^a \xi^b  + \Omega_{5\, a} \xi^a + \Omega_7) \right ) \, ,
\end{align}
and if 
\begin{align}
(d+\xi^c \iota_{R_c}) (\Omega_{3\, ab} \xi^a \xi^b  + \Omega_{5\, a} \xi^a + \Omega_7)  = \text{any two-form}\, ,
\end{align}
then $\delta S_{7,2} =0$. In detail, this means the following conditions 
\begin{align}
& (d  \Omega_{3\, ab}  + \iota_{R_a} \Omega_{5\, b} ) \xi^a \xi^b =0\, , \label{eq:obscond1}\\
& (d \Omega_{5\, a} + \iota_{R_a} \Omega_7) \xi^a=0 \label{eq:obscond2} \, .
\end{align}
On a 3-Sasaki manifold we can for example solve these conditions by
\begin{align}
\Omega_7 &= \kappa_1 (d\kappa_1)^3 = \kappa_2 (d\kappa_2)^3 = \kappa_3 (d\kappa_3)^3\, ,\\
\Omega_{5\, a} \xi^a &=-\xi^a \k_a(d\k_a)^2 = - \xi^1 \kappa_1 (d\kappa_1)^2 - \xi^2 \kappa_2 (d\kappa_2)^2 - \xi^3 \kappa_3 (d\kappa_3)^2 \,,\\
\Omega_{3\, ab} \xi^a \xi^b &= \k_a d\k_b \xi^a \xi^b \\
&=  (\xi^1)^2 \kappa_1 d\kappa_1  + (\xi^2)^2 \kappa_2 d\kappa_2 + (\xi^3)^2 \kappa_3 d\kappa_3 +\xi^1 \xi^2 \k_1d\k_2+\xi^2 \xi^1 \k_2d\k_1 +\xi^1 \xi^3\k_1d\k_3 \notag \\ 
&\quad +\xi^3 \xi^1\k_3d\k_1 +\xi^2 \xi^3 \k_2 d\k_3+\xi^3 \xi^2 \k_3 d\k_2\, . \notag
\end{align}

We can also introduce the observable
\begin{align}
S_{7,3}= \int\limits_{X} \left ({\rm Tr}  (\sigma + \Psi + F)^3 \wedge (\Omega_{1\,abc} \xi^a \xi^b \xi^c+ \Omega_{3\, ab} \xi^a \xi^b  + \Omega_{5\, a} \xi^a + \Omega_7) \right ) \, ,
\end{align}
and if 
\begin{align}
(d+\xi^c \iota_{R_c})   (\Omega_{1\,abc} \xi^a \xi^b \xi^c + \Omega_{3\, ab} \xi^a \xi^b  + \Omega_{5\, a} \xi^a + \Omega_7)  = {\rm any ~function}\, ,
\end{align}
then $\delta S_{7,3} =0$. 
This corresponds to the conditions \eqref{eq:obscond1} and \eqref{eq:obscond2} above, as well as 
\begin{align}
( d \Omega_{1\,abc}   +   \iota_{R_c}   \Omega_{3\, ab}  ) \xi^a \xi^b \xi^c=0\, .
\end{align}
To solve these conditions we can take $\Omega_{3\, ab}$, $\Omega_{5\, a}$ and $\Omega_7$ as above together with
\begin{align}
\Omega_{1\,abc} \xi^a \xi^b \xi^c &= - \left ( (\xi^1)^2 + (\xi^2)^2 + (\xi^3)^2 \right ) \xi^a \kappa_a \\
&= - \left ( (\xi^1)^2 + (\xi^2)^2 + (\xi^3)^2 \right ) \left( \xi^1 \kappa_1 + \xi^2 \kappa_2 +\xi^3 \kappa_3 \right)\, . \notag
\end{align}
These observables have  additional supersymmetry corresponding to the 3-Sasaki structure. However we do not know how to extend these additional supersymmetries to the rest of the cohomological complex and thus we do not know how to effectively use the 3-Sasaki structure in the localisation calculation.

\subsection{Factorisation}

Previously we have discussed the factorisation of the perturbative answer on toric Sasaki-Einstein manifolds. The factorisation is associated to the closed Reeb orbits and we cut our toric manifold into $S^1 \times \mathbb{C}^3$ pieces. The crucial point is that the geometrical decomposition is compatible with the decomposition at the level of  special functions. Now we would like to pose the following question: Is there an alternative decomposition related to 3-Sasaki structures? Below we look at the case of $S^7$ but we believe that our observations are valid for any 3-Sasaki manifold (at least with toric symmetry).

The seven dimensional sphere $S^7$ is a Hopf fibration in two different ways (see Figure \ref{fibrations}). Firstly, it is the standard $U(1)$-bundle over $\C P^3$. Secondly, it can be thought of as the quaternionic Hopf fibration over $S^4$, where we think of $S^4$ as $\mathbb{H}P^1$, the quaternionic analogue of $\C P^1$. This second picture is equivalent to the statement that $S^7$ is an $SU(2)$-fibration over $S^4$. 
\begin{figure}[h!]
	\centering
	\begin{tikzcd}
		S^7  \arrow[d] & S^1 \arrow[l] & & & S^7  \arrow[d] & S^3 \arrow[l]\\
		\C P ^3  & & & & S^4  & 
	\end{tikzcd}
	\caption{\textit{Hopf fibrations of $S^7$.}}\label{fibrations}
\end{figure}

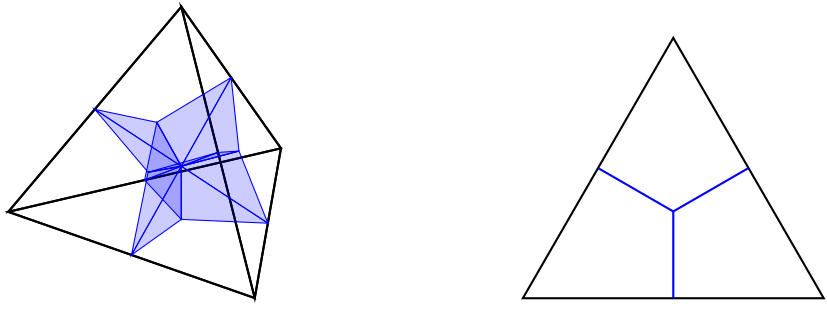
\begin{figure}[h!]
	\centering
	\tdplotsetmaincoords{60}{65}
	\begin{tikzpicture}[tdplot_main_coords, scale=4]
	\draw[thick] (0.57735,0,0)--(-0.288675,0.5,0)--(-0.288675,-0.5,0)--cycle;
	\draw[thick] (0.57735,0,0)--(-0.288675,0.5,0)--(0,0,0.816497)--cycle;
	\draw[thick] (0.57735,0,0)--(-0.288675,-0.5,0)--(0,0,0.816497)--cycle;
	\draw[thick] (-0.288675,0.5,0)--(-0.288675,-0.5,0)--(0,0,0.816497)--cycle;
	\fill[draw=blue, fill=blue, fill opacity=0.2] (0, 0,0.204124)--(0.144338,0.25,0)--(0, 0, 0)--cycle;
	\fill[draw=blue, fill=blue, fill opacity=0.2] (0, 0,0.204124)--(0.144338,0.25,0)--(0.096225, 0.166667, 0.272166)--cycle;
	\fill[draw=blue, fill=blue, fill opacity=0.2] (0, 0,0.204124)--(-0.144338, 0.25, 0.408248)--(0.096225, 0.166667, 0.272166)--cycle;
	\fill[draw=blue, fill=blue, fill opacity=0.2] (0, 0,0.204124)--(-0.288675,0,0)--(0, 0, 0)--cycle;
	\fill[draw=blue, fill=blue, fill opacity=0.2] (0, 0,0.204124)--(-0.288675,0,0)--(-0.19245, 0, 0.272166)--cycle;
	\fill[draw=blue, fill=blue, fill opacity=0.2] (0, 0,0.204124)--(-0.144338, 0.25, 0.408248)--(-0.19245, 0, 0.272166)--cycle;
	\fill[draw=blue, fill=blue, fill opacity=0.2] (0, 0,0.204124)--(0.288675, 0, 0.408248)--(0.096225, 0.166667, 0.272166)--cycle;
	\fill[draw=blue, fill=blue, fill opacity=0.2] (0, 0,0.204124)--(0.144338, -0.25,0)--(0,0,0)--cycle;
	\fill[draw=blue, fill=blue, fill opacity=0.2] (0, 0,0.204124)--(0.288675, 0, 0.408248)--(0.096225, -0.166667, 0.272166)--cycle;
	\fill[draw=blue, fill=blue, fill opacity=0.2] (0, 0,0.204124)--(0.144338, -0.25,0)--(0.096225, -0.166667, 0.272166)--cycle;
	\fill[draw=blue, fill=blue, fill opacity=0.2] (0, 0,0.204124)--(-0.144338, -0.25, 0.408248)--(0.096225, -0.166667, 0.272166)--cycle;
	\fill[draw=blue, fill=blue, fill opacity=0.2] (0, 0,0.204124)--(-0.144338, -0.25, 0.408248)--(-0.19245, 0, 0.272166)--cycle;
	\end{tikzpicture} \hspace{2,5cm}
	\tdplotsetmaincoords{0}{30}
	\begin{tikzpicture}[tdplot_main_coords, scale=4]
	\draw[thick] (0.57735,0,0)--(-0.288675,0.5,0)--(-0.288675,-0.5,0)--cycle;
	\draw[thick,draw=blue] (0,0,0)--(0.144338, 0.25,0);
	\draw[thick,draw=blue] (0,0,0)--(0.144338, -0.25,0);
	\draw[thick,draw=blue] (0,0,0)--(-0.288675,0,0);
	\end{tikzpicture}
	\caption{\textit{This figure shows the polytope that represents the $S^7$ (left) where we have performed the cut that gives four pieces of $S^1\times \C^3$. On the right one can find a projection of the polytope to one of the sides where $|z_i|^2=|z_j|^2=0$ for two specific $i\neq j$. Then the picture reduces to the 5D case.}}\label{figure2}
\end{figure}

The $U(1)$ fibration picture is telling us that $S^7$ can be represented as gluing four copies of $S^1 \times \C^3$. As we discussed in section \ref{Factor} one can factorise the quadruple sine (perturbative 1-loop answer for $S^7$) into four q-factorials and every q-factorial corresponds to
the pertubative 1-loop answer on $S^1 \times \C^3$. This decomposition is related to the closed Reeb orbits of a generic toric Reeb vector field.
Also, as illustrated in Figure \ref{figure2}, the toric diagram of the sphere can be separated into four equal pieces. Each piece contains one of the vertices where the local geometry looks like  $S^1 \times \C^3$. These vertices are where the total $T^4$ action degenerates to an $S^1$ action which is the Reeb orbit. This story is similar to what is happening in 5D, \cite{Qiu:2015rwp}.  

Now let us discuss $S^7$ as an $SU(2)$-fibration over $S^4$. This picture is related to the fact that $S^7$ is  a 3-Sasaki manifold (the metric cone of $S^7$ is $\C^4$ which is obviously hyperk\"ahler). From this point of view, $S^7$ should be glued from two copies of $S^3 \times \C^2$. Now the question is if we can see this structure at the level of special functions. Below we present a heuristic argument that this can be done and that it is consistent with the localisation picture, although there are no systematic results for the localisation of gauge theory on $S^3 \times \C^2$.

Let us analyse an alternative factorisation of the quadruple sine. Our discussion here is schematic and we ignore Bernoulli-like factors as these can be restored by studying the asymptotics of the special functions. We hope to come back to the proper treatment of this problem elsewhere.
Using the infinite product representation of multiple sine functions we have the following expression for the quadruple sine:
\begin{align}
\label{s4prod}S_4(x|\w_1,\w_2 , \w_3 ,\w_4)&= \frac{\prod\limits_{j_{1,2,3,4}=0}^{\infty}\left(x+ j_1 \w_1 + j_2 \w_2 + j_3 \w_3 + j_4 \w_4 \right)}{\prod\limits_{j_{1,2,3,4}=1}^{\infty}\left(-x+ j_1 \w_1 + j_2 \w_2 + j_3 \w_3 + j_4 \w_4 \right)}\nonumber\\&=\frac{\prod\limits_{j_{1,2,3,4}=0}^{\infty}\left(x+ j_1 \w_1 + j_2 \w_2 + j_3 \w_3 + j_4 \w_4 \right)}{\prod\limits_{j_{1,2,3,4}=0}^{\infty}\left(-x+\w_1+\w_2 +\w_3 +\w_4 + j_1 \w_1 + j_2 \w_2 + j_3 \w_3 + j_4 \w_4 \right)}\, .
\end{align}
For the double sine we can write
\begin{align}
\prod_{j_{3,4}=0}^{\infty}S_2(x+ j_3 \w_3 + j_4 \w_4 & |\w_1 , \w_2)= \frac{\prod\limits_{j_{1,2,3,4}=0}^{\infty}(x+j_1 \w_1 +j_2 \w_2 +j_3 \w_3 +j_4 \w_4)}{\prod\limits_{j_{1,2,3,4}=0}^{\infty}(-x-j_3 \w_3 -j_4 \w_4 +\w_1 +\w_2+j_1 \w_1 +j_2 \w_2)}\, ,\\
\prod_{j_{1,2}=0}^{\infty}S_2(x+ (-j_1-1) \w_1 &+ (-j_2-1) \w_2|\w_3 , \w_4)= \nn\\ &\frac{\prod\limits_{j_{1,2,3,4}=0}^{\infty}(x-\w_1 -\w_2-j_1 \w_1 -j_2 \w_2 +j_3 \w_3 +j_4 \w_4)}{\prod\limits_{j_{1,2,3,4}=0}^{\infty}(-x+\w_3 +\w_4 +j_3 \w_3 +j_4 \w_4 +\w_1 +\w_2+j_1 \w_1 +j_2 \w_2)}\, .
\end{align}
One can combine these two results and obtain the following formula:
 \begin{equation}\label{s4tos2s}
 S_4 (x| \w_1, \w_2 , \w_3 , \w_4 ) =\prod_{j_{1,2}=0}^{\infty}S_2(x+ j_1 \w_3 + j_2 \w_4|\w_1 , \w_2)S_2(x+ (-j_1-1) \w_1 + (-j_2-1) \w_2|\w_3 , \w_4)\, .
 \end{equation}
Alternatively, one can obtain the same formula (up to Bernoulli factors) using the q-factorial expansion for the multiple sine functions. However, one should then keep track of the different regions for $\omega_i$, since the concrete form of the q-factorial expansion for multiple sines depend on the sign of ${\rm Im} (\omega_i /\omega_j)$ etc.
 
The factorisation \eqref{s4tos2s} gives us some insight into the building blocks of the partition function for the sphere $S^7$. More specifically, we believe that the infinite product of double sines correspond to the perturbative part of the partition function on $S^3 \times \C^2$. This statement requires proper treatment, but here is a rough argument: 
If we look at the theory on the $S^3$ part, then the result of the superdeterminant \eqref{superdeterminant} would be a regular double sine. However, in the most general case there are additional contributions coming from an infinite tower of modes $z^m_1 z^n_2$ where $z_{1,2}$ are the coordinates on the two copies of $\C$. Thus the final result for the perturbative part of the partition function for $S^3 \times \C^2$ is of the form
\begin{equation}
Z^{\text{pert}}_{S^3\times \C^2}= \prod_{m,n=0}^{\infty}S_2(x+ m \w_3 + n \w_4|\w_1 , \w_2)\, ,\label{S3-C2-result}
\end{equation}
where $m,n$ are equivariant parameters associated with the $U(1)$ rotations in the two copies of $\C$ and $\w _{1,2}$ and $\w_{3,4}$ are the pieces of the Reeb associated with the $S^3$ and $\C ^2$ respectively. As a result the perturbative part of the partition function on $S^7$ can be written schematically as
\begin{equation}
Z^{\text{pert}}_{S^7}= \prod_{i=1}^2 Z^{\text{pert}}_{S^3\times \C^2} (a_{i1}, a_{i2}, \epsilon_{i1} , \epsilon_{i2})\, ,
\end{equation}
where there is a relation between the parameters $a_1, a_2$ associated with the $S^3$ and $\epsilon_1, \epsilon_2$ associated with the rotations in $\C^2$, analogously to \eqref{perturbatives1}. 

\begin{figure}
	\centering
\tdplotsetmaincoords{70}{115}
\begin{tikzpicture}[tdplot_main_coords, scale=4]
\draw[thick] (0.57735,0,0)--(-0.288675,0.5,0)--(-0.288675,-0.5,0)--cycle;
\draw[thick] (0.57735,0,0)--(-0.288675,0.5,0)--(0,0,0.816497)--cycle;
\draw[thick] (0.57735,0,0)--(-0.288675,-0.5,0)--(0,0,0.816497)--cycle;
\draw[thick] (-0.288675,0.5,0)--(-0.288675,-0.5,0)--(0,0,0.816497)--cycle;

\filldraw[thick , draw=black, fill=blue, fill opacity=0.2] (0.144338, -0.25,0)--(-0.288675,0,0)--(-0.144338, 0.25, 0.408248)--(0.288675,0,0.408248)--cycle;
\end{tikzpicture}
\caption{\textit{This figure shows the polytope that represents the $S^7$ where we have performed the cut that gives two pieces of $S^3\times \C^2$. }}\label{figure3}
\end{figure}
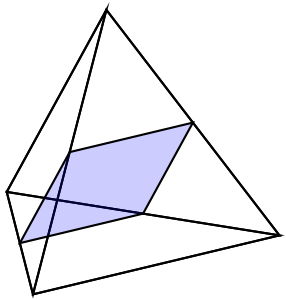

This splitting can also be seen at the level of geometry using local coordinates. The sphere can be represented by the equation $\sum_{i=1}^{4}|z_i|^2=1$, where $z_i \in \C$ and using this one can perform the following splitting:
\begin{align}
|z_1|^2 + |z_2|^2 \leq \frac{1}{2}\quad \text{and}  \quad|z_3|^2 + |z_4|^2 \geq \frac{1}{2}\, ,\\
|z_3|^2 + |z_4|^2 \leq \frac{1}{2}\quad \text{and} \quad |z_1|^2 + |z_2|^2 \geq \frac{1}{2}\, .
\textbf{•}\end{align}
Both cases can be associated to $S^3\times \C^2$ and they cover the full space. Thus they are the only contributions that we will get in the partition function as already argued above. One can also see this by cutting the polytope of $S^7$ into two pieces, see Figure \ref{figure3}. Each piece contains a pair of vertices corresponding to the $S^3$. This can be contrasted with the cut in Figure \ref{figure2} where each piece contained a single vertex.

Analogously to the Sasaki-Einstein case where the partition function localises on closed Reeb orbits, it is natural to say that for the 7-sphere it also localises on $S^3$ membranes. To make this statement more precise we need to better understand what the Nekrasov partition function for $S^3 \times \C^2$ is, and if it can be defined in principle. 

We believe that what we have suggested here for $S^7$ can be proven for any 3-Sasaki manifold with toric symmetry. Namely, that the perturbative result can be factorised into copies of the 1-loop partition function of $S^3 \times \C^2$ given by \eqref{S3-C2-result}. In any case it will require further detailed study.

\section{Summary}\label{sec:summary} 

In this paper we have constructed the maximally supersymmetric 7D Yang-Mills theory on Sasaki-Einstein manifolds. We rewrote the theory in terms of differential forms and obtained the cohomological complex which is defined for any K-contact manifold. Formally we can localise this supersymmetric theory on any Sasaki-Einstein manifold (or using its cohomological complex for any K-contact manifold). The localisation locus looks complicated and its structure is unclear to us. If we specify to the case of toric 7D Sasaki-Einstein manifolds we are able to the produce a concrete expression for the perturbative answer (expansion around zero connection)
\begin{equation}\label{xpert-summary}
Z_{X} = \int\limits_t d \sigma ~e^{- P(\sigma)} \prod_\beta S_4^{C_\mu(X)}( i \bra \sigma, \beta \ket |\vec{R})  + \dots \, ,
\end{equation}
where 
\begin{equation}
P(\sigma) = \sum\limits_{i=1}^4 a_i \Tr (\sigma^i)\, ,
\end{equation}	
with different powers of $\sigma$ corresponding to supersymmetric Yang-Mills action and 
supersymmetric Chern-Simons-like terms. The integrand $S_4^{C_\mu(X)}$ is a specific special function which is defined in terms of toric data for a given toric Sasaki-Einstein manifold.
   
The dots in \eqref{xpert-summary} stand for non-perturbative contributions which are hard to conjecture at the moment. From one side the toric manifold can be decomposed into $S^1 \times \C^3$ pieces which correspond to closed orbits for generic toric Reebs. This decomposition is consistent with the decomposition of $S_4^{C_\mu(X)}$ in terms of q-factorials which are the perturbative answers for $S^1 \times \C^3$. Thus it appears natural to assume that the full answer will be glued from copies of the Nekrasov partition function on $S^1 \times \C^3$. However we suspect that this picture is oversimplified, there may be other configurations (not associated to closed Reeb orbits) which contribute to the path integral. For example, in the case of $S^7$ we can cut the manifold into two copies of $S^3 \times \C^2$ and it looks consistent with the corresponding decomposition of perturbative 1-loop result. However we do not know what the Nekrasov partition function for $S^3 \times \C^2$ is, which should correspond to some membrane theory on the moduli space of instantons.  
       
Another important issue is about the UV completion of our 7D calculation. Typically when we look at 5D and 6D theories, they are not good QFTs as they stand. However for 5D and 6D theories the localisation calculation is well-defined and it produces sensible results. Physically these results are understood as some calculations within the appropriate UV completion. For example, the calculation for maximally supersymmetric 5D Yang-Mills corresponds to the calculation for (2,0) 6D SCFT and there exist non-trivial checks for this statement (e.g., see \cite{Kallen:2012zn}). Other localisation calculations in 5D can be related to N=1 5D SCFT and again it has been checked in some cases \cite{Jafferis:2012iv}. The maximally supersymmetric 6D Yang-Mills theory is related to the Little string theory. In the case of 7D we do not know any non-gravitational UV completion (see \cite{Losev:1997hx} for some speculations), however all our calculations are well-defined. Indeed we hope that the study of 7D theory may bring some additional insight to low dimensional theories by considering 7D theory on manifolds of the form $M_k \times M_{7-k}$ etc. 
\vspace{1cm}

 {\bf Acknowledgements}\\
 
We thank Joseph Minahan, Jian Qiu and Jacob Winding for discussions, and the anonymous referee for comments and suggestions. The research is supported in part by Vetenskapsr\r{a}det under grant \#2014-5517, by the STINT grant and by the grant ''Geometry and Physics'' from the Knut and Alice Wallenberg foundation.


\appendix 

\section{Conventions and useful identities} \label{AppConv}

Our conventions and notation follow closely those of \cite{Pestun:2007rz,Minahan:2015jta}. Here we also collect several useful identities found in e.g. the appendices of those papers.

\subsection{Notational conventions}
Throughout the paper we will denote ten-dimensional space-time indices by capital Latin letters from the middle of the alphabet, $M,N,P,Q = 0,\dots,9$. Lower case Greek letters from the middle of the alphabet will denote the seven-dimensional indices, $\mu,\nu,\lambda,\rho = 1,\dots,7$. Upper case Latin letters from the beginning of the alphabet will be used to denote the scalar field directions, $A,B,C = 0,8,9$.

Lower case Greek letters from the beginning of the alphabet, $\alpha, \beta, \gamma,\delta$, will denote spinor indices, but we will rarely write them out explicitly.

\subsection{Gamma matrices}

In $\R^{9,1}$ we take $g^{MN}$ to be the standard flat metric with mostly plus. We represent the Clifford algebra $Cl(9,1)$ by the $32 \times 32$ matrices $\gamma^M$, $M=0,\dots,9$ satisfying the anti-commutation relations
\begin{equation}
\gamma^{\{M} \gamma^{N\}} \equiv \halfs (\gamma^{M} \gamma^{N} + \gamma^{N} \gamma^{M}) = g^{MN}.
\end{equation}
The spin representation of $Spin(9,1)$ is decomposed into $S^\pm$ with eigenvalues $\pm 1$ under the chirality operator $\gamma^{11} \equiv \gamma^1\gamma^2 \cdots \gamma^9 \gamma^0$. 
The Dirac spin representation of $Spin$(9,1) is taken to be in the form 
\begin{equation}
\left(\begin{matrix} S^+ \\ S^- \end{matrix}\right),
\end{equation}
so that the $\gamma$-matrices obtain the block form
\begin{equation}
\gamma^M=\left( \begin{matrix}
0& \Gt ^M\\
\G^M & 0
\end{matrix}\right).
\end{equation}
The $16 \times 16$ Dirac $\Gamma$-matrices are taken to be real and symmetric,
\begin{align}
\tensor{\G}{^M^\alpha^\beta} = \tensor{\G}{^M^\beta^\alpha}, \qquad  \tensor{\Gt}{^M^\alpha^\beta} = \tensor{\Gt}{^M^\beta^\alpha}.
\end{align}
The anti-commutation relations for the $\G$'s are
\begin{equation}
\Gt ^ {\{ M} \G ^{N\}} \equiv \tfrac{1}{2} \left( \Gt^M \G^N +\Gt^N \G^M  \right)= g^{MN},  \quad \G ^ {\{ M} \Gt ^{N\}}\equiv \tfrac{1}{2} \left( \G^M \Gt^N +\G^N \Gt^M  \right) = g^{MN}.
\end{equation}

The anti-symmetrised products of $\G$-matrices are given by
\begin{equation}
\Gamma^{MN} \equiv \tilde\Gamma^{[M}\Gamma^{N]} \equiv \tfrac{1}{2} \left( \Gt^M \G^N -\Gt^N \G^M  \right),\qquad 
\tilde\Gamma^{MN}\equiv \Gamma^{[M}\tilde\Gamma^{N]} \equiv \tfrac{1}{2} \left( \G^M \Gt^N -\G^N \Gt^M  \right),
\end{equation}
which can be generalised to higher products of $\G$-matrices. 

From these definitions one can derive several useful relations, such as the commutation relation
\begin{equation} \label{gcommrel}
\G^M \G^{PQ}= 4g^{M[P}\G^{Q]} + \Gt^{PQ}\G^M \, ,
\end{equation}
and identities like
\begin{align} \label{gident}
&\Gt^\mu \G _\nu \Gt_\mu = -5 \Gt _\nu \, ,\nn\\
&\Gt^\mu \G _A \Gt_\mu= -7 \Gt _A \, ,\nn\\
&\Gamma^\mu \Gamma_{\nu \rho} \Gammat_{\mu} = 3\Gammat_{\nu \rho} \, ,\\
&\Gamma^\mu \Gamma_{\nu A} \Gammat_{\mu} = 5 \Gammat_{\nu A} \, ,\nn\\
&\Gamma^\mu \Gamma_{AB} \Gammat_{\mu} = 7 \Gammat_{AB}. \nn
\end{align}

A very important relation is the triality identity
\begin{equation}\label{triality}
\Gamma^M_{\alpha\beta}\Gamma_{M\gamma\delta}+\Gamma^M_{\beta\delta}\Gamma_{M\gamma\alpha}
+\Gamma^M_{\delta\alpha}\Gamma_{M\gamma\beta}=0\,.
\end{equation}
It is used to show that ten-dimensional Yang-Mills is supersymmetric, and we can also use it to show identities such as
\begin{equation} \label{eq:eGeeGx}
\eps\Gamma^M\eps\,\eps \Gamma_M\chi=0\, ,
\end{equation}
for any spinor $\chi$, where $\eps$ is a bosonic spinor. 

For our 7-dimensional theories we will define
\begin{equation}
\Lambda \equiv \G^8 \Gt^9 \G^0.
\end{equation}
It is straightforward to check that $\Lambda$ is anti-symmetric
\begin{equation}
\tensor{\Lambda}{^\alpha^\beta} = -\tensor{\Lambda}{^\beta^\alpha},
\end{equation}
and satisfies 
\begin{equation}
\La \Gt_\mu \La = - \G_\mu.
\end{equation}

We also have the following identities
\begin{align}
\La \G_{\mu\nu} &= \Gt _{\mu\nu} \La \nn\, ,\\ 
\La \G_{A \mu} &=-\Gt _{A\mu} \La \label{GaLaEps}\, ,\\
\La \G_{AB} &= \Gt _{AB} \La = - \varepsilon_{ABC}\G^C\, , \nn
\end{align}
where in the last identity $\varepsilon_{890} = +1$.

\subsection{7D spinors vs 10D spinors} \label{AppKill}

Here we give some details on the relation between Killing spinors in 7D and spinors satisfying the equation \eqref{killcond} in 10D.

In 7D we can represent the Clifford algebra $Cl(7)$ by Hermitian, purely imaginary, $8 \times 8$ matrices $\gamma_7^\mu$, $\mu=1 , \dots, 7$. Note that these matrices are also anti-symmetric.

We can write our 10D Dirac matrices $\G$ in terms of the 7D gamma matrices as follows:
\begin{align} \label{eq:10Dvs7DGamma}
&\G^\mu = 
\begin{pmatrix} 
0 & i\gamma_7^\mu\\
-i\gamma_7^\mu & 0\\
\end{pmatrix}, \mu=1,...,7, \\ 
&\G^8 = 
\begin{pmatrix} 
0 & I \\
I & 0 \\
\end{pmatrix}, \quad 
\G^9 = 
\begin{pmatrix} 
I & 0\\
0 & -I\\
\end{pmatrix}, \quad
\G^0 = 
\begin{pmatrix} 
I & 0\\
0 & I\\
\end{pmatrix}. \notag
\end{align}

Note that in this representation $\G^M = \Gt^M$, except for $\G^0 = - \Gt^0$.

Assume that we have a collection of 7D Majorana spinors $\eta_i$, which may or may not be linearly independent, satisfying the 7D Killing spinor equation
\begin{equation}
\nabla_\mu \eta_i = \alpha_{i} {\gamma_7}_\mu \eta_i\, .
\end{equation}

From these we construct a 10D Majorana-Weyl spinor 
\begin{equation}
\eps = \begin{pmatrix} \eta_i \\ \eta_j \end{pmatrix}\, .
\end{equation}
Noting that 
\begin{align}
\La \Gt_\mu \eps = 
\begin{pmatrix}
0 & -I\\
I & 0\\
\end{pmatrix}
\begin{pmatrix} 
0 & i{\gamma_7}_\mu\\
-i{\gamma_7}_\mu & 0\\
\end{pmatrix}
\begin{pmatrix} \eta_i \\ \eta_j \end{pmatrix}
= \begin{pmatrix} i{\gamma_7}_\mu\eta_i \\ i{\gamma_7}_\mu\eta_j \end{pmatrix} \, ,
\end{align}
we see that $\eps$ satisfies the generalised Killing spinor equation
\begin{equation}
\nabla_\mu \eps = \frac12 \La \Gt_\mu \eps\, ,
\end{equation}
provided that ${\alpha}_i={\alpha}_j=+\frac12 i$. The 7-dimensional manifolds with such Killing spinors are proper $G_2$-manifolds, Sasaki-Einstein manifolds, 3-Sasakian maifolds and the 7-sphere, as discussed in section \ref{sectionKilling}.
 
We next consider the vector field
\begin{equation}
v^M = \eps \G^M \eps
\end{equation}
and its restriction $v^\mu$ to the 7D space.
In terms of 7D gamma matrices we have
\begin{equation}
v^\mu = \eps \G^\mu \eps 
 = \begin{pmatrix} \eta_i & \eta_j  \end{pmatrix}
\begin{pmatrix}  0 & i\gamma_7^\mu  \\ -i\gamma_7^\mu & 0  \end{pmatrix}
\begin{pmatrix} \eta_i \\ \eta_j  \end{pmatrix}
= i\eta_i \gamma_7^\mu\eta_j -i\eta_j \gamma_7^\mu\eta_i\, .
\end{equation}

This will be zero if $\eta_i$ and $\eta_j$ are linearly dependent. For proper $G_2$-manifolds, where we only have one Killing spinor, $v^\mu$ will be identically zero. For Sasaki-Einstein, 3-Sasaki  and $S^7$, where we have at least two linearly independent Killing spinors, we can construct a non-zero $v^\mu$.

Let us now consider $v^A$, $A=8,9,0$. From \eqref{eq:10Dvs7DGamma} we get
\begin{align}
v^8 &= \eps \G^8 \eps  =  \eta_i \eta_j + \eta_j \eta_i \, ,\\
v^9 &= \eps \G^9 \eps  = \eta_i \eta_i - \eta_j \eta_j \, ,\\
v^0 &= \eps \G^{10} \eps = \eta_i \eta_i + \eta_j \eta_j \, .
\end{align}

If we have two linearly independent Killing spinors we may w.l.o.g. take them to be orthogonal and of norm-squared $1/2$.  This corresponds to the choice $v^8=v^9=0, v^0=1$, and as a consequence of \eqref{eq:eGeeGx} we then have $v^\mu v_\mu=1$.

For the proper $G_2$ case, where $\eta_j$ is necessarily proportional to $\eta_i$, we may for instance choose $v^8=0, v^9=v^0=1$. Note that since $v^\mu=0$ this is still consistent with \eqref{eq:eGeeGx}.

\section{Supersymmetry calculations} \label{AppSusy}
In this appendix we give an outline of the rather lengthy supersymmetry calculations. These calculations rely heavily on the various identities in Appendix \ref{AppConv}.

\subsection{Flat 10D case}
As a warm-up, we verify that the action \eqref{10daction}
\begin{equation}
S_{10}= \frac{1}{g_{10}^2}\int d^{10}x \Tr \left( \halfs F^{MN}F_{MN} - \Psi \G^M D_M \Psi \right) \, ,
\end{equation}
is invariant under the supersymmetry transformations \eqref{susytransflat}
\begin{align}
\delta_\epsilon A_M &= \epsilon \G_M \Psi \, ,\nn\\
\delta_\epsilon \Psi &= \halfs \G^{MN} F_{MN} \epsilon\, .
\end{align}

The variation of the first term is
\begin{align}
\deps (\halfs F_{MN}F^{MN})&= 2(\deps A_M) D_N F^{MN}  = 2(\epsilon \Gamma_M \Psi) D_N F^{MN}\, . 
\end{align}

Using the triality identity \eqref{triality} we find the variation of the second term to be
\begin{align}
\deps (- \Psi \Gamma^M D_M \Psi ) &= (\Psi \Gamma^M \Gamma^{PQ} \epsilon) D_M F_{PQ}\, .
\end{align}

Using the relations in Appendix \ref{AppConv} we can rewrite 
\begin{align}
\Gamma^M \Gamma^{PQ} &= \tfrac{1}{3} \left( \Gamma^M \Gamma^{PQ} + \Gamma^P \Gamma^{QM} + \Gamma^Q \Gamma^{MP} \right) + 2g^{M[P}\Gamma^{Q]}\, .
\end{align}
When we contract this with $D_M F_{PQ}$ the terms in the bracket cancel by the Bianchi identity and what is left cancels with the variation from the first term in the action.

\subsection{7D on-shell SUSY}

In this subsection we show that the action \eqref{7Daction}
\begin{align}
S_{7D}= \frac{1}{g_{7D}^2} \int d^{7}x \sqrt{-g} \Tr \Big( 
\halfs F^{MN}F_{MN} 
- \Psi \G^M D_M \Psi 
+8 \phi^A \phi_A
+\tfrac{3}{2} \Psi \Lambda \Psi
-2 [\phi^A,\phi^B]\phi^C \varepsilon_{ABC}
\Big)\, ,
\end{align}
is invariant under the supersymmetry transformations \eqref{susytranson}
\begin{align}
\delta _\eps  A_M &= \eps \G _M \Psi \, ,\nn \\
\delta _\eps \Psi &= \tfrac{1}{2} F_{MN}\G ^{MN}\eps + \tfrac{8}{7} \G^{\mu B} \phi _B \nabla _\mu \eps\, .
\end{align}

Finding the variation of the first term proceeds as in the flat case
\begin{align}
\deps (\halfs F_{MN}F^{MN})&= 2(\epsilon \Gamma_M \Psi) D_N F^{MN} \, .
\end{align}
For the second term we now have to take into account the additional terms in the variation of $\Psi$ and that the derivative of $\eps$ is no longer zero.
Due to this we no longer get total cancellation between the first two terms and after some calculations we end up with
\begin{align}
\deps \left( \halfs F^{MN}F_{MN} - \Psi \G^M D_M \Psi  \right) 
&=  \tfrac{3}{2} F_{\mu\nu} \eps \Gt^{\mu\nu} \La \Psi
+3 F_{A\mu} \eps \Gt^{A \mu} \La \Psi  \nn \\
&\qquad -\tfrac{9}{2}  F_{AB} \eps \Gt^{AB} \La \Psi 
-28 \phi_A \eps \Gamma^{A}\Psi \, .
\end{align}

The variation of the third term is
\begin{align}
\deps \left(8\phi^A \phi_A \right) &= 16 \phi_A \eps \Gamma^{A}\Psi \, .
\end{align}

From the fourth term we get 
\begin{align}
\deps \left( \tfrac{3}{2} \Psi \Lambda \Psi \right)
&= -\tfrac{3}{2} F_{\mu\nu}\eps\tilde\G^{\mu\nu}\La\Psi 
-3 F_{A\mu}\eps\tilde\Gamma^{A\mu }\La \Psi  \nn \\
&\qquad-\tfrac{3}{2} F_{AB}\eps\Gt^{AB}\La\Psi
+12  \phi_A \eps \G^A \Psi \, .
\end{align}

The total variation from the first four terms is thus
\begin{align}
\deps \left(  \halfs F^{MN}F_{MN} - \Psi \G^M D_M \Psi +8 \phi^A \phi_A  + \tfrac{3}{2} \Psi \Lambda \Psi \right)
&= -6 F_{AB}\eps\Gt^{AB}\La\Psi \, .
\end{align}

Since the derivatives along the compactified directions are zero we have $F_{AB}= [\phi_A,\phi_B]$ and using \eqref{GaLaEps} we find
\begin{align}
\deps \left(  \halfs F^{MN}F_{MN} - \Psi \G^M D_M \Psi +8 \phi^A \phi_A  + \tfrac{3}{2} \Psi \Lambda \Psi \right)
&= 6 [\phi_A,\phi_B] (\eps\G^{C}\Psi) \varepsilon_{ABC}\, .
\end{align}

This cancels with the variation from the fifth term
\begin{align}
\deps \left( - 2[\phi^A,\phi^B]\phi^C \varepsilon_{ABC} \right) 
&= -6 [\phi^A,\phi^B] (\eps \G^C \Psi) \varepsilon_{ABC}\, ,
\end{align}
and thus the total supersymmetry variation is zero.

\subsection{7D off-shell SUSY}\label{offshell}

Here we sketch the arguments for showing that the off-shell supersymmetry transformations \eqref{susytransoff} square to symmetries of the theory. Details can be found in \cite{Minahan:2015jta}.

It is straightforward to check that \cite{Minahan:2015jta}
\begin{align}
\deps^2 A_\mu  &= -v^\nu F_{\nu\mu} + [D_\mu,v^A\phi_A] \, ,\\
\deps^2 \phi_A &= -v^\nu D_\nu\phi_I-[v^B\phi_B,\phi_A]- 4 \eps \Gt_{IJ}\La \eps\ \phi^J  \label{phit}\, .
\end{align}
In both expressions, the first term is (the negative of) the Lie derivative along the (Reeb) vector field $v$ and the second term is a gauge transformation. The third term in the second equation is an $R$-symmetry transformation. Thus the two variations close up to symmetries of the action. 

The $\deps^2$-calculations for the fields $\Psi$ and $K^m$ are rather lengthy, but the final result is \cite{Minahan:2015jta}
\begin{align}
\deps^2 \Psi &= -v^M D_M \Psi - \frac{1}{4}(\nabla_{[\mu}v_{\nu]})\G^{\mu\nu}\Psi -(\eps\Gt^{AB}\La\eps)\G_{AB}\Psi \label{deltapsi}\, ,\\
\deps^2 K^m & =-v^M D_M K^m-(\nu^{[m}\G^\mu\nabla_\mu \nu^{n]})K_n+\tfrac{3}{2} (\nu^{[m}\La\nu^{n]})K_n \label{deltak}\, .
\end{align}
In \eqref{deltapsi} one can identify the Lie derivative, the gauge transformation and the last term is consistent with the $R$-symmetry. The transformation of the auxiliary field in \eqref{deltak} gives the Lie derivative and an internal $SO(7)$ transformation amongst the $K^m$.

Therefore, the operator $\delta^2_\eps$ can be seen as
\begin{equation}
\label{deltasquare}
\delta^2_\eps = - \L_v - G- R - S\, ,
\end{equation}
where $G$ is a gauge transformation, $R$ is the $R$-symmetry and $S$ is the $SO(7)$ transformation. 

\section{Multiple sine functions}\label{AppSine}

In this appendix we collect some facts about the special functions appearing in the paper. We refer the reader to \cite{Narukawa,Winding:2016wpw} for a thorough discussion.

Let $\wvec = (\w_1, \dots, \w_r) \in \C^r$ be such that $\re (\w_i)>0$\footnote{More generally, we can allow the $\w_i\in \C$ to all lie on the same side of some straight line passing through the origin of the complex plane.}. Then we can define the multiple zeta function as
\begin{equation}
\zeta_r(s,x|\wvec)= \sum_{\vec{n} \in \Z^r_{\geq 0}}\frac{1}{(x+\vec{n} \cdot \wvec)^s}\, ,
\end{equation}
for $x \in \C$ and $\re s > r$. This function can be analytically continued to $s\in \C$ and it is holomorphic at $s=0$. We use it to define the multiple Gamma function via
\begin{equation}
\G_r(x|\wvec) = \exp \left(  \frac{\partial}{\partial s} \zeta_r(s,x|\wvec) \bigg|_{s=0} \right) \, .
\end{equation}

Then the multiple sine function can be defined to be
\begin{equation}
\label{sinefunction}
S_r (x|\wvec ) = \Gamma_r (x| \wvec ) ^{-1} \Gamma_r (|\wvec| -x |\wvec )^{(-1)^r}\, ,
\end{equation}
where $|\wvec| = \sum_{i=1}^r \w_i$.

These special functions are important because they turn up in localisation calculations on spheres in various dimensions, see \cite{Pestun:2016jze}.

We will mostly work with the infinite product representations of these functions. These are given by
\begin{equation}
\Gamma_r (x|\wvec) = \prod_{\vec{n} \in \Z^r_{\geq 0}} (\vec{n}\cdot \wvec + x)^{-1}\, ,
\end{equation}
and
\begin{equation}
S_r (x| \wvec) = \prod_{\vec{n} \in \Z^r_{\geq 0}}(\vec{n}\cdot \wvec + x)\prod_{\vec{n} \in \Z^r_{> 0}}(\vec{n}\cdot \wvec - x)^{(-1)^{r-1}} \, .
\end{equation}

Following \cite{Winding:2016wpw} we will now generalise these functions.

Let us begin by defining what is meant by a `good cone'. This concept was first introduced by Lerman \cite{Lerman}, but here we instead give an equivalent definition \cite{Winding:2016wpw}:
A cone $C$ of dimension $r$ is \emph{good} if for every codimension $k$ face, the $k\times r$ matrix $[u_{i_1}\ldots u_{i_k}]$ can be completed into an $SL_r (\Z)$ matrix, where the $u_i$ are the associated $k$ normals.

Following \cite{Winding:2016wpw} we define $n^k$ to be the generating vector for the ray $k$ of the cone. By the goodness condition one can always find an $n^k\in \Z^r$ such that 
\begin{equation}
\text{det}[n^k, v_1^k,\ldots, v^k_{r-1}]=1\, ,
\end{equation}
where $ v_1^k,\ldots, v^k_{r-1}$ is a set of normal vectors to the ray $k$. Note that the choice of $n^k$ is not unique. Then we can define elements of $SL_r(\Z)$ associated to the generators to be
\begin{equation}
\tilde{K}_k = [n^k, v_1^k,\ldots, v^k_{r-1}]^{-1}\, , 
\end{equation}
and they depend on the choice of $n^k$. 

A simple example of a good cone is the moment map cone of $S^7$ which is $\R^4_{\geq 0}$. Then one choice of the four elements of $SL_4(\Z)$ is $\tilde{K}_1 = \text{Id}_4$, and the rest are cyclic permutations of the rows of this matrix, i.e. $(\tilde{K}_2)_{1,\cdot}= (0,1,0,0)$ and so on.

We can now define the generalised multiple sines \cite{Winding:2016wpw}.
Let $C$ be a good cone of dimension $r$. Assume that $\wvec \in \C^r$ is such that there exist a $k\in \C^\ast$ such that $\re(k\w)\in \left(\check{C}\right)^o$. Here $\left(\check{C}\right)^o$ denotes the interior of the dual cone. The generalised multiple sine function associated to $C$ is defined as
\begin{equation}
\label{gsinefunction}
S_r^C (x| \wvec ) = \Gamma_r^C (x| \wvec )^{-1} \Gamma_r^{C^o}(-x |\wvec )^{(-1)^r}\, ,
\end{equation}
where $\Gamma _r ^C $ is the generalised multiple gamma function associated to $C$
\begin{equation}
\Gamma_r^C (x|\wvec) = \exp \left( \frac{\partial}{\partial s}\zeta _r ^C(s,x|\wvec)\bigg|_{s=0}\right)\, ,
\end{equation}
which is defined using the generalised multiple zeta function 
\begin{equation}
\zeta _r ^C(s,x|\wvec)= \sum_{\vec{n}\in C\cap\Z^r}\frac{1}{(x+\vec{n} \cdot \wvec)^s}\, .
\end{equation}
Note that for the cone $C= \R_{\geq 0}^r$ the generalised multiple sine function \eqref{gsinefunction} reduces to the usual multiple sine function\eqref{sinefunction}.


\providecommand{\href}[2]{#2}\begingroup\raggedright


\providecommand{\href}[2]{#2}\begingroup\raggedright\endgroup

\endgroup

\end{document}